\documentclass[]{article}

\usepackage[a4paper, total={6in, 8in}]{geometry}

\usepackage{authblk}

\usepackage{graphics} 
\usepackage{epsfig} 
\usepackage{amsmath} 
\usepackage{amssymb}  
\usepackage{amsthm}	
\usepackage{cite}

\usepackage[caption=false]{subfig}

\usepackage[table]{xcolor}

\usepackage{capt-of}
\usepackage{mathtools}

\usepackage{booktabs}
\usepackage{multirow}

\newcommand{\R}{\mathbb{R}}

\newcommand{\event}{\lozenge}
\newcommand{\always}{\square}
\newcommand{\bconj}{\bigwedge}
\newcommand{\bdisj}{\bigvee}

\newcommand{\rmm}[1]{\left< \mathrm{#1} \right>}

\newcommand{\CS}[1]{\texttt{CS}$_#1$}

\newcommand{\sign}{\mathrm{sign}}

\newtheorem{rem}{Remark}
\newtheorem{thm}{Theorem}
\newtheorem{prop}{Proposition}
\newtheorem{cor}{Corollary}
\newtheorem{exmp}{Example}[section]
\newtheorem{definition}{Definition}[section]
\newtheorem{prob}{Problem}[section]
\newtheorem{assum}{Assumption}[section]

\usepackage{footnote}
\makesavenoteenv{tabular}
\makesavenoteenv{table}

\usepackage{scrextend}

\begin{document}

\title{Formal controller synthesis for hybrid systems using genetic programming.} 

\author{Cees F. Verdier and Manuel Mazo Jr.} 
\date{\vspace{-5ex}}

\affil{Delft Center for Systems and Control, Delft University of Technology, The Netherlands (e-mail: c.f.verdier@tudelft.nl, m.mazo@tudelft.nl)}
\maketitle

\begin{center}
\thanks{Supported by NWO Domain TTW under the CADUSY project \#13852. }

\end{center}

\begin{abstract}
This paper proposes a framework for automatic formal controller synthesis for general hybrid systems with a subset of safety and reachability specifications. The framework uses genetic programming to automatically co-synthesize controllers and candidate Lyapunov-like functions. These candidate Lyapunov-like functions are used to formally verify the control specification, and their correctness is proven using a Satisfiability Modulo Theories solver. The advantages of this approach are: no restriction is made to polynomial systems, the synthesized controllers are expressed as compact expressions, and no explicit solution structure has to be specified beforehand. We demonstrate the effectiveness of the proposed framework in several case studies, including nonpolynomial systems, sampled-data systems, systems with bounded uncertainties, switched systems, and systems with jumps. 
\end{abstract}

\section{Introduction}

With advances in automation and control, specifications beyond the traditional stability requirements become increasingly more relevant. These more advanced specifications can be formulated in temporal logics \cite{Katoen2008}, where the combination of reachability and safety is a simple example. Moreover, these controllers are often implemented in embedded hardware, adding complexities such as sampled data and quantization, which results in intrinsically hybrid systems. Formal synthesis for general hybrid systems with temporal logic specifications lacks a constructive controller design, making it an intricate process.

Nevertheless, in recent years, tools have been developed for formal control synthesis for this class of problems. Most of these methods fit into one of three main paradigms: synthesis by means of 1) finite (bi-)simulation abstractions \cite{Belta2017, Tabuada2009}, 2) online optimization-based methods \cite{Belta2019}, and 3) control Lyapunov and/or barrier functions \cite{Arstein1983, Wieland2007}.

The first paradigm relies on discretization of the state space and therefore suffers from the curse of dimensionality, resulting in controllers taking the form of enormous look-up tables, which complicates their implementation \cite{Zapreev2018}. Control approaches using this paradigm include \cite{Fainekos2009, Reissig2017,Liu2013, Habets2006, Girard2010}, whereas tools implementing this paradigm include PESSOA \cite{Mazo2010}, SCOTS \cite{Rungger2016}, CoSyMa \cite{Mouelhi2013} and ROCS \cite{Li2018}. Optimization-based methods typically employ model-predictive control to optimize a cost function related to the temporal logic specification, see e.g. \cite{Raman2015, Sadraddini2018} and the survey \cite{Belta2019}, hence typically require online optimization.

The certificate paradigm infers temporal properties indirectly by means of certificate functions, e.g. Lyapunov functions and barrier certificates \cite{Prajna2004}. Similarly, \textit{control} certificate functions, such as control Lyapunov functions \cite{Arstein1983,Cairano2014, Sanfelice2016} and control barrier functions \cite{Wieland2007}, can be used to design a control input such that the closed-loop system satisfies the desired properties. Using these (control) certificate functions or combinations thereof, (a subset of) temporal properties can be inferred indirectly \cite{Romdlony2016, Ames2016,Xu2015, Srinivasan2018,Lindemann2019, Garg2019}. For general hybrid systems, \cite{Han2020,Han2020ltl} recently proposed a set of sufficient conditions for certificate functions for temporal logic operators. To go beyond single temporal operators, the temporal logic formula can be decomposed into a sequence of sub formulae, resulting in a sequence of certificate functions that impose the full specification, see e.g. \cite{Dimitrova2014,Wongpiromsarn2016,Han2020ltl,Bisoffi2018,Bisoffi2020}. Regardless, synthesizing these functions for general hybrid systems is nontrivial. Many synthesis approaches rely on sum of squares approaches or counterexample-guided inductive synthesis (CEGIS) methods. The former, see e.g. \cite{Papachristodoulou2002, Tan2004, Prajna2007}, relies on polynomial systems and/or solutions.  However, even if a polynomial closed-loop system is asymptotically stable, this does not imply that there exists a polynomial Lyapunov function, as shown in \cite{Ahmadi2011}. On the other hand, CEGIS approaches, including \cite{Ravanbakhsh2015,Ravanbakhsh2017, Kapinski2014,Ahmed2020}, synthesize controllers and/or certificate functions by iteratively proposing and verifying candidate solutions. The verification typically utilizes a Satisfiability Modulo Theories (SMT) solver \cite{Barrett2009}; a numerically sound tool capable of verifying whether a first-order logic formula is satisfied or not. These CEGIS approaches do not restrict to polynomials, but typically require the user to provide a template solution. In recent work, neural networks have been used within a CEGIS framework for verification and/or formal controller synthesis \cite{Abate2020,Chang2019}. However, neural networks are not as insightful and compact as analytic expressions.  

In this work, we also use the paradigm of certificate functions, but do not constrain ourselves to polynomial dynamics and solutions and do not require solution templates. To achieve this, we propose a CEGIS framework in which we combine genetic programming (GP) \cite{Koza1992} with SMT solvers to automatically synthesize correct-by-design controllers for hybrid systems with a combination of simple safety and reachability requirements. This is done by co-synthesizing a controller and Lyapunov barrier-like function which are provably correct. Genetic programming is an evolutionary algorithm capable of optimizing solution structures consisting of pre-defined elementary building blocks. The user is thus not required to supply an explicitly parametrized structure beforehand, such as e.g. a fixed-order polynomial. We use a variant of grammar guided genetic programming (GGGP) \cite{McKay2010,Verdier2018}, which employs grammars to constrain the search space. The ability to automatically search over the space of solution structures is particularly useful when no solution exists for a certain parametrization, as the algorithm explores other structures automatically. However, the drawback is a method that is not complete, i.e. it might not return a solution in a fixed number of iterations, even if such solution exists. The resulting controllers are closed-form compact expressions, as opposed to the solutions from abstraction-based and optimization-based methods.

This paper extends upon the existing literature on CEGIS-based synthesis of (control) certificate functions by considering general hybrid systems and automatically evolving the solution structures.
The proposed method differs from previous work on GP for Lyapunov function synthesis in e.g. \cite{Grosman2009,Mcgough2010} and/or controller synthesis in e.g. \cite{Fleming2002,Koza2003,Sekaj2007,Kadlic2014,diveev2015,Reichensdorfer2017} in that the proposed method provides formal guarantees. The main contribution of this work is to synthesize controllers in the form of analytic expressions by extending our previous work \cite{Verdier2017,Verdier2018} to general hybrid systems modeled as jump-flow systems with differential and difference inclusions. 

\section{Preliminaries}

Let $\R_{\geq 0} = \{x\in \R| x\geq 0 \}$ and $\mathbb{N} = \left\{ 0,1,2 \dots \right\}$. Given a set $D\subset \R^n$, we denote the boundary and the interior with $\partial D$ and $int(D)$ respectively. The image of set $D$ under $f$ is denoted by $f[D]$. A vector in $\R^n$ comprising of only zeros or ones is denoted as $\mathbf{0}_n$ and $\mathbf{1}_n$ respectively. Finally, the Euclidean norm is denoted by $\| \cdot \|$. A table with the most important symbols can be found in Appendix \ref{app:LOS}.

In this paper we adopt the jump-flow system formalism from \cite{Goebel2012}. 
We briefly recall the following definitions:

\begin{definition}[Hybrid time domains {\cite[Def. 2.3]{Goebel2012}}]
A subset $E\subset \R_{\geq 0} \times \mathbb{N}$ is a compact hybrid time domain if $E = \bigcup_{j=0}^{J-1} ([t_j,t_{j+1}],j)$ with $0 = t_0 \leq t_1 \dots \ \leq t_J$. It is a hybrid time domain if $E \cap ([0,T] \times \{0,1,\dots J \})$ is a compact hybrid domain for all $(T,J) \in E$.
\end{definition}

Given a hybrid time domain $E$ and a given $j \in \mathbb{N}$, let us denote a time interval $T^j := \{ t \mid (t,j) \in E \}$, $E_{\leq (T,J)} := E \cap ( [0,T] \times [0,J])$, and $E_{\geq (T,J)} := E\backslash ( [0,T) \times [0,J))$.
\begin{definition}[Hybrid arc {\cite[Def. 2.4]{Goebel2012}}] A function $\phi : E \rightarrow \R^n$ is a hybrid arc if $E$ is a hybrid time domain and if for each $j \in \mathbb{N}$ the function $t \mapsto \phi(t,j)$ is locally absolutely continuous on the interval $T^j$.
\end{definition}

\begin{figure}[t]%
\centering
\subfloat[Flow set $C$, jump set $D$ and a solution $\phi(t,j)$.]{\includegraphics[width = 0.20\textwidth ]{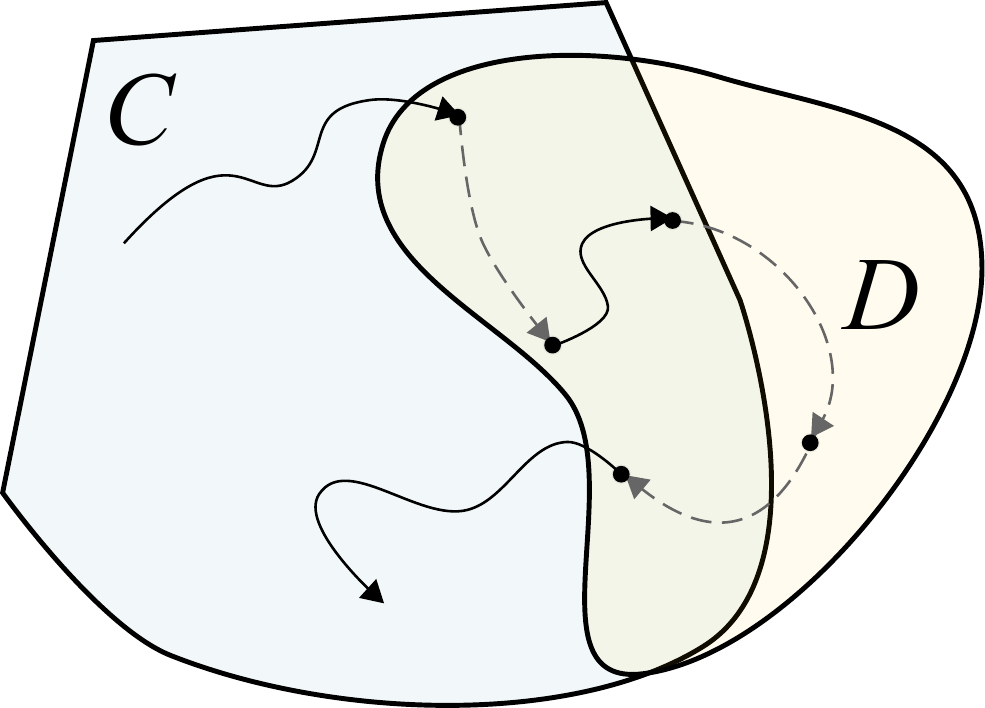} \label{fig:dyn}}  \hspace{2cm}
\subfloat[Safe set $S$, initial set $I$ and goal set $O$.]{\includegraphics[width = 0.20\textwidth ]{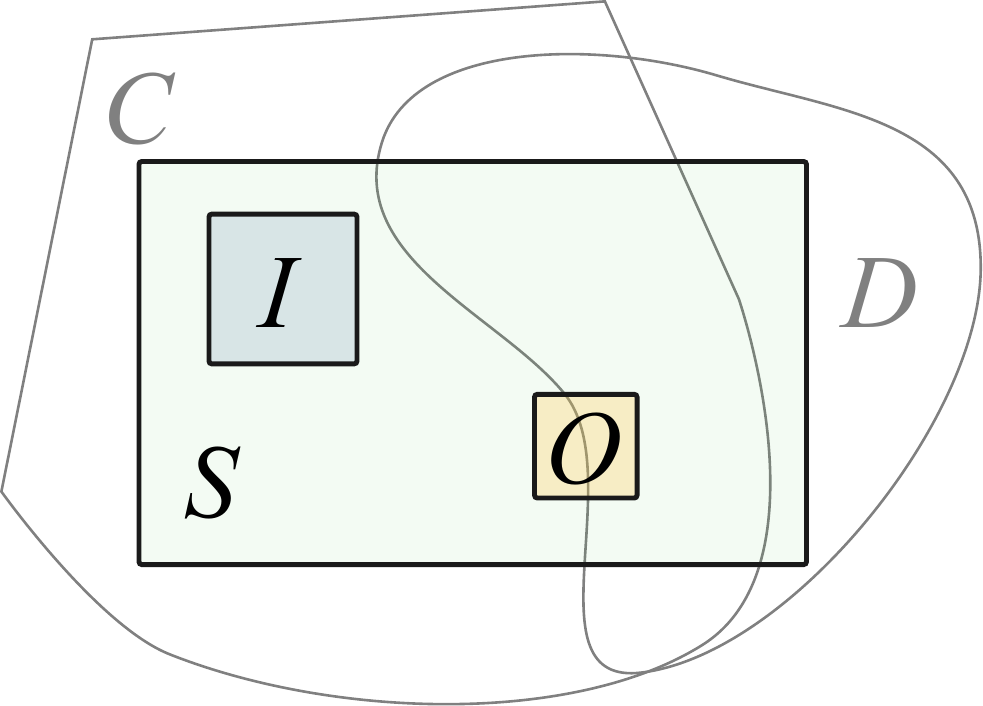} \label{fig:spec}}
\caption{Example sets}
\label{fig:sets}
\end{figure}

\begin{definition}[Hybrid system {\cite[\S 2.1]{Goebel2012}}]
\label{def:hybridsystem}
A hybrid system $\mathcal{H}$ is defined as a tuple $(C,F,D,G)$, where $C \subset \R^n$ is the flow set, set-valued function $F:C \rightrightarrows \R^n$ the flow map, $D \subset \R^n$ the jump set, and set-valued function $G: D \rightrightarrows \R^n$ the jump map. 
\end{definition}
Given a set-valued function $M: \R^m \rightrightarrows \R^n$, we denote its domain with $\mathrm{dom} M$, defined as
$\mathrm{dom} M := \{ x \in \R^m \mid M(x) \neq \emptyset \}.$
We assume the considered hybrid systems satisfies the so-called hybrid basic conditions \cite{Goebel2012}:
\begin{assum}[Hybrid basic conditions {\cite[Ass. 6.5]{Goebel2012}}]
\label{assump:basicHybrid}
~
\begin{enumerate}
\item $C$ and $D$ are closed subsets of $\R^n$.
\item $F:\R^n\rightrightarrows \R^n$ is outer semicontinuous and locally bounded relative to $C$, $C \subset \mathrm{dom} F$, and $F(x)$ is convex for every $x \in C$.
\item $G:\R^n \rightrightarrows \R^n$ is outer semicontinuous and locally bounded relative to $D$, and $D \subset \mathrm{dom} G$. 
\end{enumerate}
\end{assum}
For the definition of outer semicontinuity and local boundedness for set-valued mappings we refer to Definition 5.9 and 5.14 in \cite{Goebel2012}. Under the hybrid basic conditions, solutions to the hybrid system are defined as follows: 
\begin{definition}[Solution to a hybrid system {\cite[\S 6.2.1]{Goebel2012}}]
A hybrid arc $\phi : E \rightarrow \R^n$ is a solution to a hybrid system $\mathcal{H}$ if $\phi(0,0) \in C \cup D$ and %
\begin{itemize}
\item 
$\forall j \in \mathbb{N}$ and almost all $t \in T^j: \phi(t,j) \in C,~
\dot{\phi}(t,j) \in F(\phi(t,j))$.
\item $\forall (t,j) \in \left\{ (t,j) \in E \mid (t,j+1) \in E\right\}: \phi(t,j) \in D,~\phi(t,j+1) \in G(\phi(t,j))$.
\end{itemize}
\end{definition}
Figure \ref{fig:dyn} illustrates an example of the flow and jump sets $C$ and $D$, and a solution $\phi(t,j)$. A solution $\phi:E \rightarrow \R^n$ is complete if its domain $E$ is unbounded. Furthermore, it is Zeno if it is complete and $\sup \{ t \in \R_{\geq 0} \mid \exists j \in \mathbb{N}: (t,j) \in E \} < \infty$, i.e. an infinite number of jumps within a finite time interval. A solution $\phi$ to $\mathcal{H}$ is maximal if there exists no solution $\psi$ to $\mathcal{H}$ such that $\mathrm{dom} \phi \subset \mathrm{dom} \psi$ and $\phi(t,j) = \psi(t,j)$ for all $(t,j) \in \mathrm{dom} \phi$. Finally, we denote $\mathcal{S}_\mathcal{H}(I)$ as the set of all maximal solutions $\phi:E \rightarrow \R^n$ to $\mathcal{H}$ with $\phi(0,0) \in I$. 

\section{Problem definition}
Let us consider a state space $\R^n$, input space $\R^m$ and output space $\R^l$. Given a flow set $C$, jump set $D$, and open-loop flow map $F_\mathrm{ol}: \R^n \times \R^m \rightrightarrows \R^n$, open-loop jump map $G_\mathrm{ol}: \R^n \times \R^m \rightrightarrows \R^n$ and output map $h: \R^n \rightarrow \R^l$, the goal of this paper is to design a static output-feedback controller $\kappa: \R^l \rightarrow \R^m$, resulting in a closed-loop hybrid system $\mathcal{H}_\mathrm{cl} = (C,F,D,G)$ with $F(s) = F_\mathrm{ol}(s,\kappa \circ h(s))$ and $G(s) = G_\mathrm{ol}(s,\kappa \circ h(s))$. These controllers are expressed as analytic expressions and are therefore referred to as \textit{analytic controllers}. The controllers are designed for specifications in terms of safety w.r.t. a safe set and reachability w.r.t. a goal set for solutions starting in an initial set. We consider compact safe sets $S \subset C \cup D$, compact initial sets $I \subset S$ and compact goal sets $O \subset S$, which can be represented as
\begin{equation}
Y = \left\{ s \in C \cup D ~\middle| ~\bigwedge_{i=1}\nolimits^{i_Y} b_{Y,i}(s) \leq 0 \right\}
\label{eq:set_standardform}
\end{equation}
for $Y \in \{S,I,O\}$, with $i_Y>0$ and $b_{Y,i}: \R^n \rightarrow \R$ for $i \in \{1, \dots i_Y \}$. The main reason for choosing bounded sets is for numerical and practical reasons within the automatic synthesis and verification. An example of these sets is shown in Figure \ref{fig:spec}. Now given the sets $(S,I,O)$, and solutions $\phi: E \rightarrow \R^n$, consider the following closed-loop specifications\footnote{Representing the specifications as signal temporal logic \cite{maler2004}, we have $\varphi_\mathtt{CS_1}=\varphi_I \wedge \varphi_S \mathcal{U}_{[0,\infty)} \varphi_O,$ and $\varphi_\mathtt{CS_2}= \varphi_I \wedge \always_{[0,\infty)} \varphi_S \wedge \event_{[0,\infty)} \always_{[0,\infty)} \varphi_O$, where $\varphi_Y: \R^n \rightarrow \mathbb{B}$ denotes the set membership predicate w.r.t. set $Y$.}:
\begin{enumerate}
\item[\CS{1}] \textit{Reach while stay (RWS)}: all maximal solutions $\phi$ to $\mathcal{H}_\mathrm{cl}$ starting from the initial set $I$ eventually reach the goal set $O$, while staying within the safe set $S$:
\end{enumerate}
\begin{equation}
\begin{array}{r}
\forall \phi \in \mathcal{S}_{\mathcal{H}_\mathrm{cl}}(I) ,\exists (T,J) \in E, \forall (t,j) \in E_{\leq (T,J)} : \\
 \phi(t,j) \in S \wedge \phi(T,J) \in O.
\end{array}
\label{eq:RWS}
\end{equation}
\begin{enumerate}
\item[\CS{2}] \textit{Reach and stay while stay (RSWS)}: all maximal solutions $\phi$ to $\mathcal{H}_\mathrm{cl}$ starting from the initial set $I$ eventually reach and stay in the goal set $O$, while always staying within the safe set $S$:
\end{enumerate}
\begin{equation}
\begin{array}{r}
\forall \phi \in \mathcal{S}_{\mathcal{H}_\mathrm{cl}}(I),\exists (T,J) \in E,\forall (t,j) \in E, \\
\forall (a,b) \in E_{\geq (T,J)}: \phi(t,j) \in S \wedge \phi(a,b) \in O.
 \end{array}
 \label{eq:RSWS}
\end{equation}
Note that satisfying specification \CS{1} or \CS{2} does not preclude that complete solutions of system $\mathcal{H}_\mathrm{cl}$ exhibit Zeno behavior. Corollaries \ref{cor:noZeno} and \ref{cor:noZeno2} will address this issue. Moreover, note that specification \CS{2} does not impose that solutions should stay in $O$ after the first time instant it enters $O$, but rather that for each solution there exists a time instant $(T,J) \in E$ after which it stays in $O$. With the definition of the system and specifications, we are ready to define the following problem: 
\begin{prob}
\label{prob:1}
Given a specification \CS{1} or \CS{2} w.r.t. compact sets $(S,I,O)$ and the open-loop system $(C,F_\mathrm{ol},D,G_\mathrm{ol},h)$, synthesize an analytic controller $\kappa: \R^l \rightarrow \R^m$ such that the closed-loop system satisfies the specification.
\end{prob}

Next to synthesizing the controller for the flow/jump map, in some applications it is desired to design the flow set and jump set as part of the hybrid controller, for example in the synthesis of a supervisory controller that determines which controller mode should be active.  Consider open-loop flow and jump sets $C_\mathrm{ol}, ~D_\mathrm{ol}$ dependent on the controller $\kappa:\R^l \rightarrow \R^{m}$ such that $C = C_\mathrm{ol}(\kappa \circ h(x)),~D = D_\mathrm{ol}(\kappa \circ h(x))$. This yields the following variation of Problem \ref{prob:1}:
\begin{prob}
\label{prob:2}
Given a specification \CS{1} or \CS{2} w.r.t. compact sets $(S,I,O)$ and the open-loop system $(C_\mathrm{ol},F_\mathrm{ol},D_\mathrm{ol},G_\mathrm{ol},h)$, synthesize an analytic controller $\kappa: \R^l \rightarrow \R^{m}$ such that the closed-loop system satisfies the specification.
\end{prob}%

\section{Lyapunov barrier functions}
\label{sec:LBF}

\begin{figure*}[t]%
\centering
\subfloat[Sublevel set $A$.]{\includegraphics[width = 0.18\textwidth ]{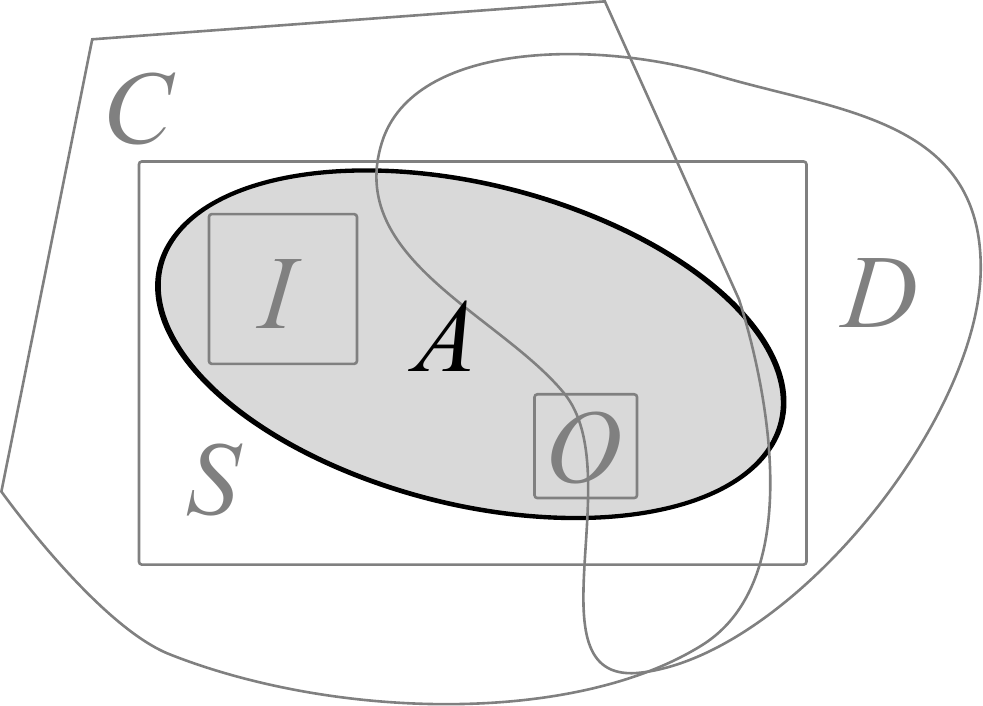} \label{fig:A}} \hfill%
\subfloat[Set $A^*:= A\backslash O$.]{\includegraphics[width = 0.18\textwidth ]{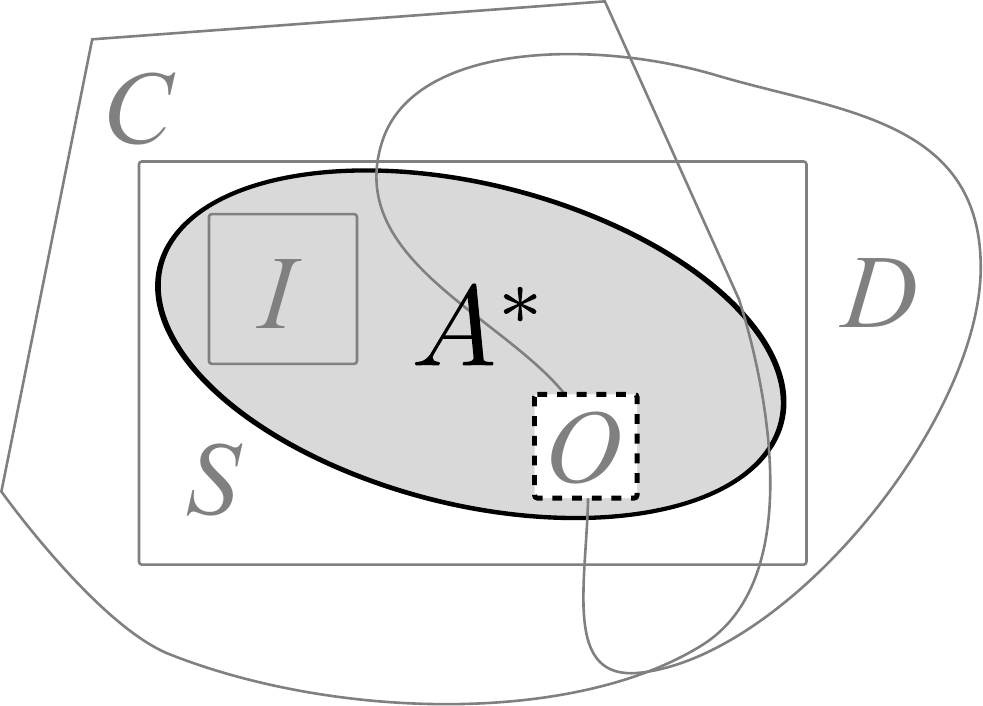} \label{fig:Astar}} \hfill
\subfloat[Set $A^*_C:= A^* \cap C$.]{\includegraphics[width = 0.18\textwidth ]{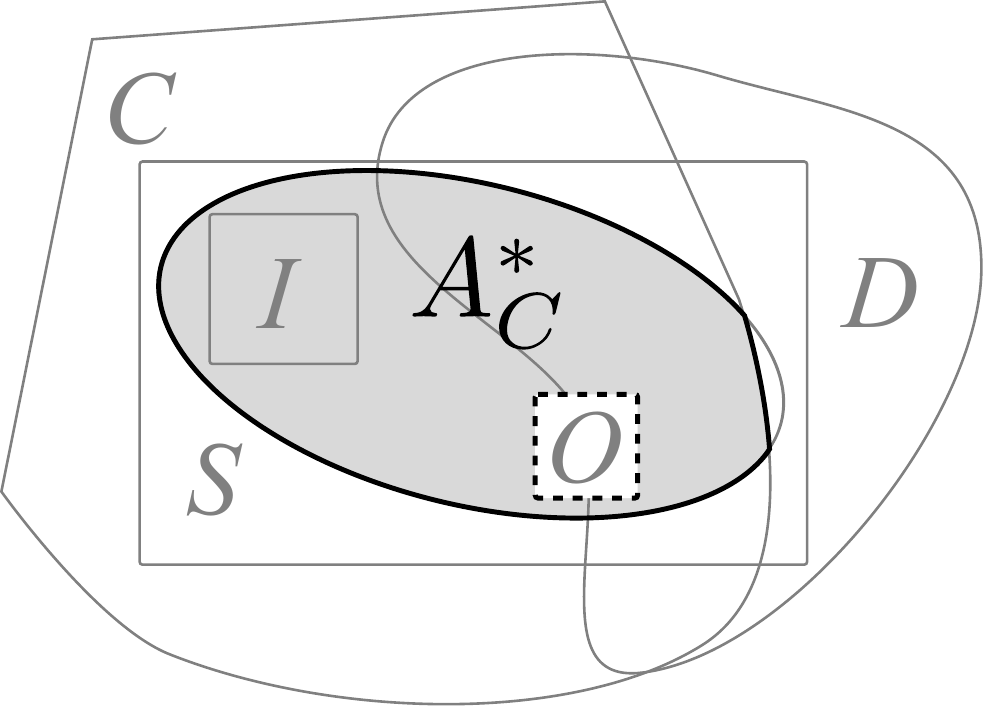} \label{fig:AstarC}} \hfill
\subfloat[Set $A^*_D:= A^* \cap D$]{\includegraphics[width = 0.18\textwidth ]{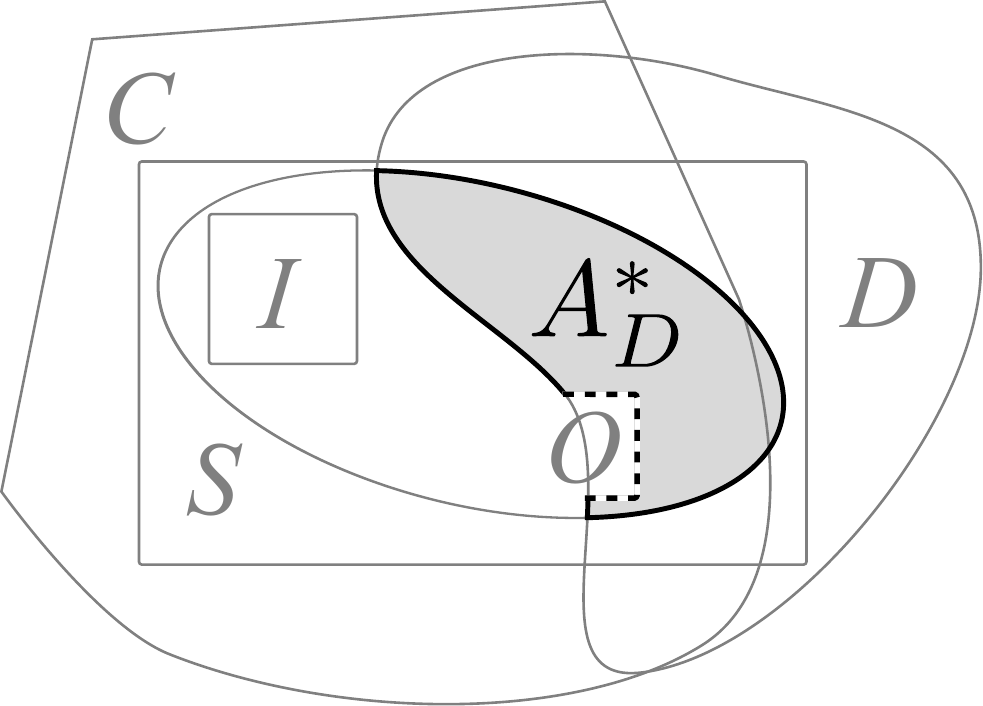} \label{fig:AstarD}}%
\caption{Example of the sublevel set $A$ and sets $A^*$, $A^*_C$ and $A^*_D$.}
\label{fig:setsA}
\end{figure*}

In this paper we verify specification \CS{1} or \CS{2} by means of a Lyapunov barrier function (LBF) which is co-synthesized with the controller. In this section, we present an LBF in Definition \ref{def:V} and present relaxations thereof in Section \ref{sec:relax}. The proofs of the technical results are presented in Appendix \ref{app:proofs}. Definition \ref{def:V} is similar to Lyapunov and/or barrier functions for hybrid systems as proposed in \cite{Prajna2004,Han2020,Han2020ltl}, to which we consider slight modifications for the purpose of automatic synthesis. In particular, the LBF conditions are posed as nonlinear inequalities over the reals, which are in general not decidable. Therefore, the synthesis and verification rely on $\delta$-decidability instead, in which a perturbed version of the inequalities are used, see \cite{Gao2010}. As a consequence, the LBF conditions are proposed with this constraint in mind. With a similar reasoning, we assume that the goal set $O$ has a nonempty interior. As remarked earlier, with the purpose of using SMT solvers to verify the conditions, we assume that the sets $(S,I,O)$ are compact. Consider the following assumption:
\begin{assum}[Specification sets assumption]
The compact sets $(S,I,O)$ can be expressed in the form \eqref{eq:set_standardform}, $S \subset C \cup D$, $O,I \subseteq int(S)$ and $int(O) \neq \emptyset$.
\end{assum}

\begin{rem}[Existence of solutions]
\label{rem:existence}
Under the hybrid basic conditions on $\mathcal{H}_\mathrm{cl}$, it follows from Proposition 6.10 in \cite{Goebel2012} that for all $s \in I \subseteq int(S) \subset C \cup D$, there exists a nontrivial solution $\phi$ to $\mathcal{H}_\mathrm{cl}$ with $\phi(0,0) = s$. 
\end{rem}
\begin{definition}[Lyapunov barrier function]
\label{def:V}A function $V \in \mathcal{C}^1(S, \R)$ is a Lyapunov barrier function w.r.t. the compact sets $(S,I,O)$ and system $\mathcal{H}_\mathrm{cl}$, if there exist $\gamma_\mathrm{c}, \gamma_\mathrm{d} > 0$ such that
\begin{subequations}
\begin{align}
\forall s \in I: ~V(s) \leq 0, \label{eq:conI} \\
\forall s \in \partial S: ~V(s) > 0, \label{eq:condS0} \\
 \forall s \in A^*_{D}: ~G(s) \subseteq S, \label{eq:conVjump} \\
\forall s \in A^*_C, \forall f \in F(s): \langle \nabla V(s), f \rangle \leq -\gamma_\mathrm{c}, \label{eq:conVd} \\
\forall s \in A^*_{D}, \forall g \in G(s):
 V(g)-V(s) \leq -\gamma_\mathrm{d}, \label{eq:conVj0} 
\end{align}\label{eq:conV0}%
\end{subequations}%
where $A^* := A \backslash O$, for $Y \in \{C, D\}$, $A^*_Y:= A^* \cap Y$ and
\begin{align}%
A &:= \{ s\in S \mid V(s) \leq 0\}. \label{eq:defA}%
\end{align}%
\end{definition}%
The sublevel set $A$ and its subsets are illustrated in Figure \ref{fig:setsA}. Set $A$ is in some sense similar to both a basin of attraction of $O$ and a forward invariant set (up until the goal set is reached), and it contains the initial set $I$, as by condition \eqref{eq:conI}.
The basin of attraction-like nature stems from \eqref{eq:conVd} and \eqref{eq:conVj0}, which impose that during flow and jumps the value of the LBF decreases. The forward invariant-like nature of $A$ stems from conditions \eqref{eq:condS0} and \eqref{eq:conVjump}, which impose that during flow and jumps, solutions cannot leave the safe set $S$ and due to the decrease need to remain within $A$. Finally, it can be proven that these properties are sufficient to imply that trajectories eventually have to enter $O$ while staying in $S$, as is formalized in Theorem \ref{thm:1} and its proof.
\begin{thm}[Reach while stay]
\label{thm:1}
Given the closed-loop system $\mathcal{H}_\mathrm{cl}$, if there exists an LBF $V$ w.r.t. compact sets $(S,I,O)$, then the closed-loop system satisfies \eqref{eq:RWS}.
\end{thm}

\begin{figure}%
\centering
\subfloat[]{\includegraphics[width = 0.15\textwidth ]{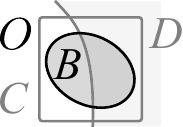} \label{fig:B}} ~ ~~~~~%
\subfloat[]{\includegraphics[width = 0.12\textwidth ]{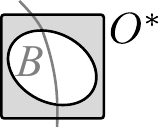} \label{fig:Gstar}} ~~
\subfloat[]{\includegraphics[width = 0.135\textwidth ]{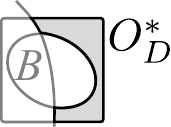} \label{fig:GstarCD}}
\caption{Given the sets from Figure \ref{fig:A}, an example of (a) the sublevel set $B$, (b) set $O^*:= O \backslash int(B)$, and (c) set $O^*_D := O^* \cap D$.}
\label{fig:setsG}
\end{figure}

The LBF implies that states within the sublevel set $A$ enter the goal set $O$ in finite time. However, it does not imply that trajectories entering $O$ stay there, nor stay in the safe set. The following corollary to Theorem \ref{thm:1} gives sufficient conditions such that specification \CS{2} is enforced.
\begin{cor}[Reach and stay while stay]
\label{col:rsws}
Given a closed-loop system $\mathcal{H}_\mathrm{cl}$ and LBF $V$ w.r.t. compact sets $(S,I,O)$, if $\exists \beta \in \R$ such that $V$ additionally satisfies
\begin{subequations}
\begin{align}
 \forall s \in O^*_{D}: G(s) \subseteq S, \label{eq:conVjumpadd0}\\
\forall s \in O^*_C, \forall f \in F(s): \langle \nabla V(s), f \rangle \leq -\gamma_\mathrm{c}, \label{eq:conVd2} \\
\forall s \in O^*_{D}, \forall g \in G(s):
 V(g) - V(s) \leq -\gamma_\mathrm{d}, \label{eq:conVj20} \\
\forall s \in \partial O : ~V(s) > \beta, \label{eq:condG0} \\
\forall s \in B \cap D: ~G(s) \subseteq B, \label{eq:conVjumpadd}
\end{align}%
 \label{eq:addV0}%
\end{subequations}%
where $B:= \{s \in O \mid V(s) \leq \beta \}$, $O^* = O \backslash int(B)$ and for $Y \in \{C, D\}$, $O^*_Y =O^* \cap Y$, then the closed-loop system $\mathcal{H}_\mathrm{cl}$ satisfies \eqref{eq:RSWS}.
\end{cor}
The sublevel set $B$ and some subsets are illustrated in Figure \ref{fig:setsG}. Set $B$ is a forward invariant set inside the interior of $O$ and all maximum solutions starting in $I$ enter this set within finite time. Intuitively, \eqref{eq:conVjumpadd0}-\eqref{eq:conVj20} extend the set on which conditions \eqref{eq:conVjump}-\eqref{eq:conVj0} hold to include $O \backslash int(B)$. Condition \eqref{eq:condG0} and \eqref{eq:conVjumpadd0} render $B$ forward invariant, similarly to the role of conditions \eqref{eq:condS0} and \eqref{eq:conVjump} w.r.t $S$. Together, they imply that solutions enter a forward invariant subset of $O$.

Specification \CS{2} reasons over maximal solutions, but it does not exclude the possibility of Zeno behavior, as shown in the following example:
\begin{exmp}[Zeno behavior]
Consider a hybrid system with $G(s,u) = 0$ and $D = \{0\}$, which admits Zeno solutions, as each jump can be followed by another jump. Now for a goal set such that $D \subseteq O$, the existence of an LBF satisfying the conditions \eqref{eq:addV0} is not contradicted by $G$, $D$ and $O$. Therefore, an LBF satisfying Corollary \ref{col:rsws} is not sufficient to exclude the admittance of Zeno solutions.
\end{exmp}
The next corollary to Theorem \ref{thm:1} establishes a sufficient condition on $V$ such that the maximal solutions are non-Zeno. We provide no proof for this result, as it is analogous to the proof of Corollary \ref{cor:noZeno2} in the next Section.
\begin{cor}[Zeno-free solutions]
\label{cor:noZeno}
Given a closed-loop system $\mathcal{H}_\mathrm{cl}$, LBF $V$ w.r.t. compact sets $(S,I,O)$, and there exists a $\beta \in \R$ such that $V$ satisfies \eqref{eq:addV}, if $B \cap D = \emptyset$, all solutions $\phi \in \mathcal{S}_{\mathcal{H}_\mathrm{cl}}(I)$ are non-Zeno.
\end{cor}

\section{Relaxations}
\label{sec:relax}

\begin{figure*}[t]%
\centering
\subfloat[Continuous state $\phi_\mathrm{x}(t,j)$.]{\includegraphics[scale =0.8]{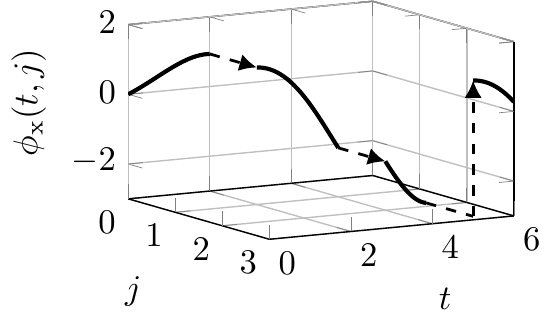} \label{fig:sx}} \hfill%
\subfloat[Discrete state $\phi_\mathrm{q}(t,j)$.]{\includegraphics[scale =0.8]{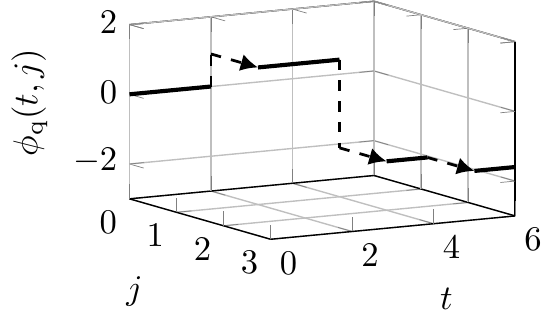} \label{fig:sq}} \hfill%
\subfloat[Timer state $\phi_\mathrm{t}(t,j)$.]{\includegraphics[scale =0.8]{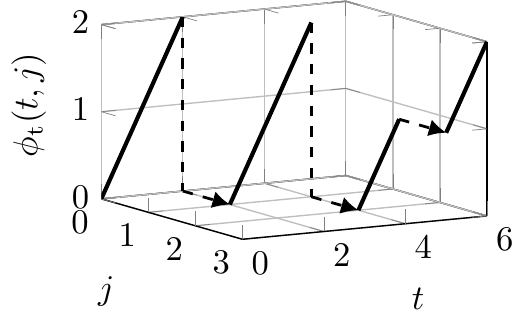} \label{fig:st}} \hfill%
\caption{Example of the evolution of the different types of states for a sampled-data system with the sampled state as discrete state. The system is subjected to $D_\mathrm{s} = \{ s \in \R^3 \mid s_\mathrm{x} \leq -3\}$, $D_\mathrm{t} = \R^2 \times \{2 \}$, $G_\mathrm{s}(s) = (1, s_\mathrm{q}, s_\mathrm{t} )$, $G_\mathrm{t}(s) = (s_\mathrm{x}, s_\mathrm{x}, 0)$, resulting in timer state-induced jumps $j \in \{1,2\}$ and system state-induced jump $j=3$.}
\label{fig:states}
\end{figure*}%

Without limiting the class of considered systems, we partition the system states into three types, allowing for relaxed LBF conditions. Given a solution $\phi(t,j)$, we distinguish continuous states $\phi_\mathrm{x}(t,j)$, discrete states $\phi_\mathrm{q}(t,j)$ and timer states $\phi_\mathrm{t}(t,j)$. The continuous states can change during both flow and jumps, whereas the discrete states can only change during jumps. The timer states increase at a constant rate during flow and each timer state $\phi_{\mathrm{t},i}(t,j)$ is reset after $\eta_i$ seconds. Now, $\phi(t,j)$ is partitioned as
\begin{align*}
\phi(t,j) &= (\phi_\mathrm{x}(t,j), \phi_\mathrm{q} (t,j),\phi_\mathrm{t}(t,j)),\\
\phi_\mathrm{x}(t,j) &\in \mathcal{X}\subseteq \R^{n_\mathrm{x}}, ~
\phi_\mathrm{q}(t,j) \in \mathcal{Q} \subseteq \R^{n_\mathrm{q}},\\
\phi_\mathrm{t} (t,j)& \in \mathcal{T} := \Pi_{i=1}^{n_\mathrm{t}}[0, \eta_i], ~\eta_i > 0.
\end{align*}
Here $C \cup D \subseteq \mathcal{X} \times \mathcal{Q} \times \mathcal{T} \subseteq \R^n$ and $n = n_\mathrm{x} + n_\mathrm{q} + n_\mathrm{t}$. Similarly, a point $s \in \R^n$ is partitioned as $s = (s_\mathrm{x},s_\mathrm{q},s_\mathrm{t})$. The timer reset motivates the distinction in two types of jumps: a timer jump if $ \phi(t,j) \in D_\mathrm{t}$ is induced by the timer resets, and a system jump if $\phi(t,j) \in D_\mathrm{s}$ is induced by the system states $(\phi_\mathrm{x}, \phi_\mathrm{q})$. During system jumps, the timer states remain constant, whereas during timer jumps, the timer state that triggered the jump is reset to zero. This yields the following system structure:
\begin{align*}
F_\mathrm{ol}(s,u) = \begin{pmatrix}
F_\mathrm{ol,x}(s,u),
\mathbf{0}_{n_\mathrm{q}}, \mathbf{1}_{n_\mathrm{t}}
\end{pmatrix},\\
G_\mathrm{ol}(s,u) = \left\{\begin{array}{ll} 
G_\mathrm{ol,s}(s,u), & \text{if } s \in D_\mathrm{s}\backslash D_\mathrm{t},\\
G_\mathrm{ol,t}(s,u), & \text{if } s \in D_\mathrm{t} \backslash D_\mathrm{s},\\
G_\mathrm{ol,s}(s,u) \cup G_\mathrm{ol,t}(s,u),  & \text{if } s \in D_\mathrm{s} \cap D_\mathrm{t} ,
 \end{array} \right. \\ 
 G_\mathrm{ol,s}(s,\!u) = \! \left( \!
G_\mathrm{ol,s}^\mathrm{xq}(s,\!u),
s_\mathrm{t}
 \! \right) \!,
G_\mathrm{ol,t}(s,\!u) = \! \left( \!
G_\mathrm{ol,t}^\mathrm{xq}(s,\!u), 0 \!
\right) \!, \\
 D_\mathrm{s} \subseteq \mathcal{X} \times \mathcal{Q} \times \mathcal{T}, ~~~
D_\mathrm{t} \subseteq \bigcup\nolimits_{i=1}^{n_\mathrm{t}} D_{\mathrm{t},i},\\
D_{\mathrm{t},i} \subseteq \mathcal{X} \times \mathcal{Q} \times \Pi_{k=1}^{i-1}[0,\eta_k] \times \{\eta_i\} \times \Pi_{k=i+1}^{n_\mathrm{t}} [0,\eta_k] \\
\mathrm{reset}(c) = (\mathrm{reset}_1(c_1), ~\dots, ~\mathrm{reset}_{n_\mathrm{t}}(c_{n_\mathrm{t}})),\\
 \mathrm{reset}_i(c_i)= \left\{
\begin{array}{ll}
0 & \text{ if } c_i = \eta_i,\\
c_i & \text{ otherwise,} 
\end{array}
\right.
 \end{align*}
where $F_\mathrm{ol,x}: \R^{n} \times \R^m \rightrightarrows\R^{n_\mathrm{x}}$ denotes the flow map for the continuous states and $G_\mathrm{ol,s}^\mathrm{xq},G_\mathrm{ol,t}^\mathrm{xq}: \R^n\times \R^m \rightrightarrows \R^{n_\mathrm{x} + n_\mathrm{q}}$ are the jump maps for the continuous and discrete states, triggered by the system states and timer state. 
The jump set of the entire system is given by $D = D_\mathrm{s} \cup D_\mathrm{t}$. The distinction between continuous, discrete and timer states and their respective behavior is illustrated in Figure \ref{fig:states}. In the remainder we use the notation $G_\mathrm{s}(s) =G_\mathrm{ol,s}(s,\kappa \circ h (s) )$ and $G_\mathrm{t}(s) = G_\mathrm{ol,t}(s,\kappa \circ h (s) )$ for the jump maps of the closed-loop system.

Examples of states that could be modeled as discrete states include logic states, discrete states, and sampled states for sampled-data systems. The timer state can be used to model the sample update of sampled-data systems. 

\begin{rem}[Absence of state types]
We allow the possibility for $n_\mathrm{x}$, $n_\mathrm{q}$, $n_\mathrm{t}$ to be zero, i.e. the absence of continuous, discrete or timer states. Subsequently, with abuse of notation, we define for the corresponding `non-existing' space $\R^0$ such that $A \times \R^0 := A$. Note that the object $\R^0$ is not equal to the empty set, as $A \times \emptyset = \emptyset$.
\end{rem}

For the three types of states, we assume that the safe set, initial set and goal set satisfy the following assumption, which helps to further relax the conditions on the candidate LBF.
\begin{assum}[Specification sets assumption revised]
\label{ass:sets}
Given compact sets $S_\mathrm{x} \subseteq \mathcal{X}$, $I_\mathrm{x}\subset int(S_\mathrm{x})$, $O_\mathrm{x}\subset int(S_\mathrm{x})$, $S_\mathrm{q} \subseteq \mathcal{Q}$, and $ O_\mathrm{q}\subseteq S_\mathrm{q}$, the compact safe, initial and goal sets $(S,I,O)$ can be expressed as in the form in \eqref{eq:set_standardform} and are defined such that:
\begin{enumerate}
\item $S:= S_\mathrm{x} \times S_\mathrm{q} \times \mathcal{T} \subseteq C \cup D$. 
\item $I \subseteq  I_\mathrm{x} \times S_\mathrm{q}  \times \mathcal{T} \subset S.$
\item $O := O_\mathrm{x} \times O_\mathrm{q} \times \mathcal{T}\subset S$ and $int(O_\mathrm{x}) \neq \emptyset. $
\end{enumerate}
\label{def:SIG}
\end{assum}
Here $S_\mathrm{x}$, $I_\mathrm{x}$ and $O_\mathrm{x}$ are the safe, initial and goal set of the continuous states and $S_\mathrm{q}$ and $O_\mathrm{q}$ the safe and goal set of the discrete states. Note that by definition the entire timer state space is considered to be in the safe and goal set.

\begin{rem}[Existence of solutions, revisited] Analogous to Remark \ref{rem:existence}, for all $s \in I \subseteq (int(S_\mathrm{x})\times S_\mathrm{q} \times \mathcal{T} )\subset C \cup D$, there exists a nontrivial solution $\phi$ to $\mathcal{H}_\mathrm{cl}$ with $\phi(0,0) = s$. 
\end{rem}

The explicit division between continuous, discrete and timer states allows for relaxations on the conditions on the candidate LBF, as well as to Corollary \ref{col:rsws} and \ref{cor:noZeno}. The proofs are presented in the Appendix \ref{app:proofs}. 
\begin{prop}[Sufficient conditions for RWS]
\label{thm:2}
Given the closed-loop system $\mathcal{H}_\mathrm{cl}$ and compact sets $(S,I,O)$ satisfying Assumption \ref{ass:sets}, if there exists a candidate LBF $V$ that satisfies \eqref{eq:conI}, \eqref{eq:conVjump}, \eqref{eq:conVd} and 
\begin{subequations}
\begin{align}
\forall s \in \partial S_\mathrm{x} \times S_\mathrm{q} \times \mathcal{T} : ~V(s) > 0, \label{eq:condS} \\
\forall s \in A^*_{D_\mathrm{s}}, \forall g_\mathrm{s} \in G_\mathrm{s}(s):
 V(g_\mathrm{s})-V(s) \leq -\gamma_\mathrm{d}, \label{eq:conVj} \\
 \forall s \in A^*_{D_\mathrm{t}}, \forall g_\mathrm{t} \in G_\mathrm{t}(s):
 V(g_\mathrm{t})-V(s) \leq 0,
 \label{eq:conVjc} %
\end{align}\label{eq:conV}%
\end{subequations}%
where for $Y \in \{D_\mathrm{s}, D_\mathrm{t}\}$, $A^*_Y:= A^* \cap Y$,
then the closed-loop system satisfies \eqref{eq:RWS}.
\end{prop}

Compared to the original LBF, it is sufficient if $V(s) >0$ holds only at the boundaries of the safe set of the continuous states, i.e. $\partial S_\mathrm{x} \times S_\mathrm{q} \times \mathcal{T}$, as during flow the discrete states and timer states cannot escape the safe set. Furthermore, due to persistent flowing and systems jumps, there is no need for the decrease during timer jumps in \eqref{eq:conVjc}. 

\begin{cor}[Sufficient conditions for RSWS]
\label{col:rsws2}
Given a closed-loop system $\mathcal{H}_\mathrm{cl}$, compact sets $(S,I,O)$ that satisfy Assumption \ref{ass:sets}, and a candidate LBF $V$ satisfying all conditions in Proposition \ref{thm:2}, if $\exists \beta \in \R$ such that $V$ additionally satisfies \eqref{eq:conVjumpadd0}, \eqref{eq:conVd2}, \eqref{eq:conVjumpadd} and
\begin{subequations}
\begin{align}
\forall s \in O^*_{D_\mathrm{s}}, \forall g_\mathrm{s} \in G_\mathrm{s}(s):
 V(g_\mathrm{s}) - V(s) \leq -\gamma_\mathrm{d}, \label{eq:conVj2} \\
 \forall s \in O^*_{D_\mathrm{t}}, \forall g_\mathrm{t} \in G_\mathrm{t}(s):
 V(g_\mathrm{t}) - V(s) \leq 0, \label{eq:conVj2c} \\
 \forall s \in \partial O_\mathrm{x} \times O_\mathrm{q} \times \mathcal{T} : ~V(s) > \beta, \label{eq:condG}%
\end{align}\label{eq:addV}%
\end{subequations}%
where for $Y \in \{D_\mathrm{s}, D_\mathrm{t}\}$, $O^*_Y =O^* \cap Y$, then the closed-loop system $\mathcal{H}_\mathrm{cl}$ satisfies \eqref{eq:RSWS}.
\end{cor}

\begin{cor}[Zeno-free solutions]
\label{cor:noZeno2}
Given a closed-loop system $\mathcal{H}_\mathrm{cl}$, a candidate LBF $V$ w.r.t. compact sets $(S,I,O)$ satisfying all conditions in Corollary \ref{col:rsws2}, if $B \cap D_\mathrm{s} = \emptyset$, all solutions $\phi \in \mathcal{S}_{\mathcal{H}_\mathrm{cl}}(I)$ are non-Zeno.
\end{cor}

Similar to the Lyapunov relaxations for hybrid inclusions in \cite[\S 3.3]{Goebel2012}, we can relax the LBF conditions further, if we have persistent jumping or persistent flowing. In this paper we only consider the latter. 

\begin{assum}[Restricted jumps]
\label{ass:jumps}
All jumps cannot be followed by additional jumps, i.e. $
\forall s \in S \cap D: G(s) \notin D$.
\end{assum}
Maximal solutions to systems that satisfy this assumption are intrinsically subjected to persistent flowing and therefore no decrease along $V$ for every jump is required:
\begin{cor}[Sufficient LBF conditions: persistent flow]
\label{cor:persflow}
Given a closed-loop system $\mathcal{H}_\mathrm{cl}$ which satisfies Assumption \ref{ass:jumps}, Theorem \ref{thm:1}, Proposition \ref{thm:2} and Corollaries \ref{col:rsws} and \ref{col:rsws2} hold with respect to $\gamma_\mathrm{d} = 0$. 
\end{cor}

\section{Automatic synthesis}
In the previous sections we derived conditions on a candidate LBF to infer specification \CS{1} or \CS{2}. In the remainder of this paper, we propose a framework to co-synthesize a controller $\kappa$ and LBF $V$. The synthesis method uses GP to propose candidate solutions, which are subsequently formally verified using an SMT solver. If a candidate solution is disproved to be a solution, the SMT solver provides a counterexample which is then used to refine the candidate solutions.

\subsection{Genetic programming}
\label{sec:GP}

In order to co-design a controller and an LBF, i.e. the tuple $(V,\kappa)$, we employ genetic programming: an evolutionary algorithm which sets itself apart in its capability to synthesize entire expressions, rather than optimizing parameters in a predefined structure. That is, given a set of elementary building blocks, the structure of the function can be modified, e.g. a polynomial can increase or decrease in order.

In GP, candidate solutions, also referred to as individuals, have two types of representation, namely the phenotype: in our case the tuple $(V,\kappa)$ expressed as a tuple of (analytic) expressions; and the genotype: an encoding of the phenotype in a form that allows for easy manipulation. This manipulation is done using so-called genetic operators, that e.g. change the genotype of an individual, or recombines multiple genotypes. Given an individual, a metric on how well the objective is achieved is captured in a fitness function. The algorithm is initialized with a randomly generated population of individuals. Each individual is scored using the fitness function. Related to the fitness, individuals are selected to undergo genetic operations, which result in a population of new individuals. This cycle is repeated until a satisfactory individual is found or a maximum number of generations (cycles) is met. 

\begin{figure*}[t]%
\centering
\subfloat[Grammar.]{\includegraphics[scale =0.8]{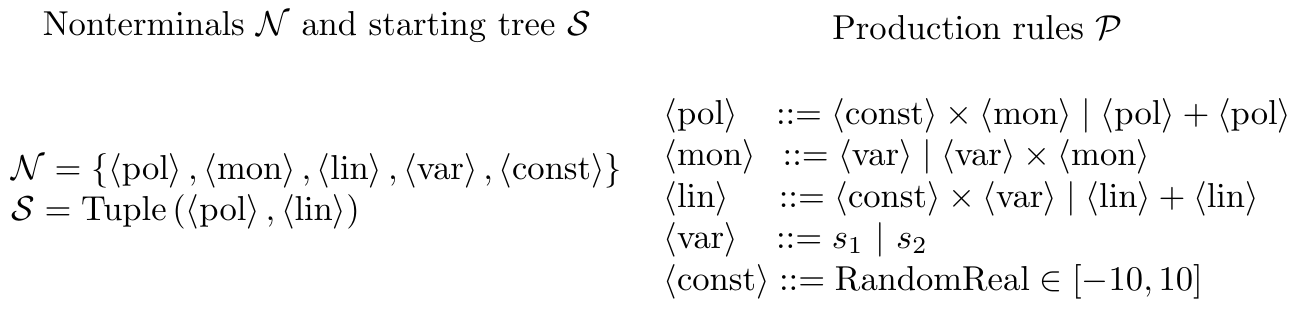} \label{fig:gram}} \hfill
\subfloat[Fully expanded genotype.]{\includegraphics[width = 0.25\textwidth ]{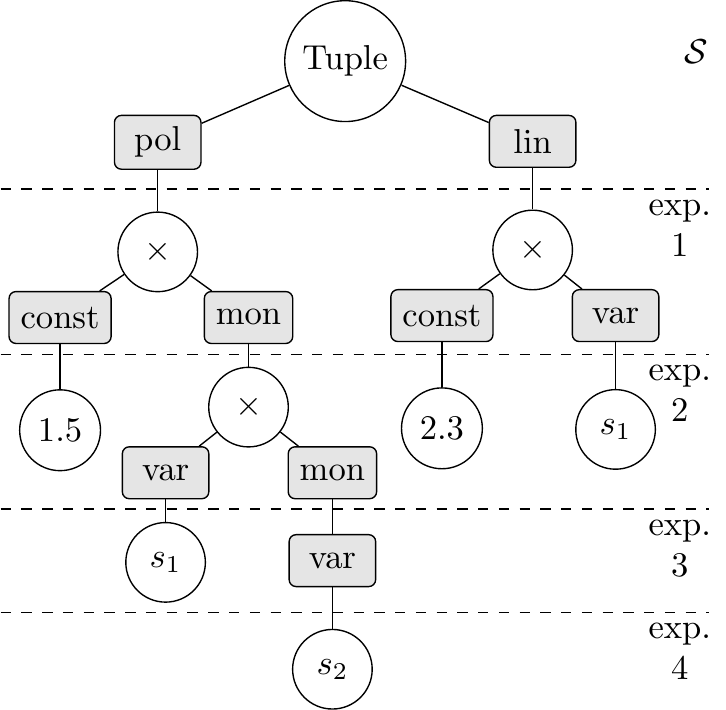} \label{fig:geno}} \hfill
\caption{Example of a grammar to synthesize a typle $(V,\kappa)$, where $V$ is polynomial and $\kappa$ is linear, and a genotype adhering to the grammar. The corresponding phenotypes is $(1.5 s_1s_2, 2.3s_1)$ }
\label{fig:grammar_examp}
\end{figure*}

We use grammar-guided genetic programming (GGGP) \cite{Verdier2017,Verdier2018}, which imposes that the genotype adheres to a certain grammar in Backus-Naur form (BNF) \cite{Backus1963}. The BNF grammar is defined by the tuple $(\mathcal{N}, \mathcal{S},\mathcal{P})$, where $\mathcal{N}$ denotes a set of nonterminals, $\mathcal{S} \in \mathcal{N}$ is a starting tree, and $\mathcal{P}$ are the production rules. An Example of a grammar $(\mathcal{N}, \mathcal{S},\mathcal{P})$ for a tuple $(V,\kappa)$ is shown in Figure \ref{fig:gram}. Given the grammar, a genotype is constructed as follows. The starting tree forms an initial expression tree. In this tree, all nonterminals are expanded by picking corresponding production rules from $\mathcal{P}$ and placing these rules under the nonterminals. For the new tree, the nonterminals in the leaf nodes are again expanded and this process is repeated, until no leaf nodes contain a nonterminal. To prevent an infinite tree depth, it is possible to pre-define a tree depth after which no recurrent production rules in $\mathcal{P}$ are used. A fully expanded genotype is shown in Figure \ref{fig:geno}. Finally, to transform the genotype into the phenotype, first all nonterminals are removed from the genotype by replacing each nonterminal node by its underlying node. Given this form without nonterminals, a phenotype is obtained by rewriting the resulting expression tree as an analytic expression. The phenotype corresponding to the genotype in Figure \ref{fig:geno} is $(1.5 s_1 s_2, 2.3 s_1)$.

We use tournament selection \cite{Koza1992} as selection method, in which a fixed number of individuals are randomly chosen from the population, and the individual with the highest fitness is returned as the selected individual. In case multiple individuals have the same fitness, secondary fitness measures (see Section \ref{sec:secfit}) are used to break the tie. The used genetic operators are crossover and mutation. In crossover, two individuals are selected and two random subtrees with the same nonterminal as root are interchanged. In mutation, a random subtree is interchanged with a randomly grown subtree with the same nonterminal. Note that the resulting trees both adhere to the same grammar as before. Finally, each generation, the constants within the evolved structure are optimized using Covariance Matrix Adaptation Evolution Strategy (CMA-ES) \cite{hansen2001}. More specifically, we use the variant sep-CMA-ES, due to its linear space and time complexity \cite{Ros2008}. 

\subsection{Verification and counterexample generation}
\label{sec:verification}
To verify formally verify the LBF conditions, we use the SMT solver dReal \cite{Gao2013}, which is able to verify nonlinear inequalities over the reals. As these are in the most general case not decidable, dReal implements a $\delta$-complete decision procedure \cite{Gao2010}, i.e. it determines whether a first order logic formula is unsatisfiable (unsat) or if the $\delta$-weakening is satisfiable ($\delta$-sat). The $\delta$-weakening can be seen as a perturbed version of the original inequality, that renders the decision process decidable. This makes it possible to formally prove whether a logic formula is not satisfied, or a perturbation is satisfied. By using the SMT solver to prove that the negation of the original logic formula is unsatisfiable, we obtain a proof of the satisfiability of the original formula. Note that unsat and $\delta$-sat are not mutually exclusive. If there is an overlap, dReal can return either case. This issue is addressed in Remark \ref{rem:dsatrobustness}. In case a formula is $\delta$-sat, dReal provides a domain in which the formula is $\delta$-sat. From this domain we can sample states that are counterexamples where the inequality is (close to be) violated.

\subsection{Fitness}
\label{sec:fitness}
The evolutionary search is driven by the fitness function. In this section we elaborate on how the fitness function is constructed. Based on the inequalities in Proposition \ref{thm:2} and Corollary \ref{col:rsws2}, we employ both testing and verification techniques to assign a fitness value to a candidate solution. Given an inequality over a set, the testing is done on a finite subset of the original infinite set. This test provides us with a quality measure of candidate solutions, and thus provides a search direction for the genetic evolution. The verification method uses the SMT solver to determine a boolean answer to whether the inequality is satisfied over the entire set.

The LBF conditions can be expressed as a propositional formula $\varphi$ in the standard form:
\begin{equation}
\varphi:= \forall x \in X: \left( \bconj_{i=1}\nolimits^{k} \left( \bdisj\nolimits_{j=1}^{l_i}f_{i,j}(x) \leq 0 \right) \right),
\label{eq:standard}
\end{equation}
where $f_{i,j}: \R^n \rightarrow \R$. The standard form of the conditions in Proposition \ref{thm:2} and Corollary \ref{col:rsws2} can be found in Appendix \ref{app:standard}. Given the propositional logic formula $\varphi$ in standard form, we formulate for a point $x \in X$ a satisfaction measure $\rho_\varphi: \R^n \rightarrow \R$ as:
\begin{equation}
\rho_\varphi(x) = \max_{i \in \{1, \dots, k\}} \left( \min_{j \in \{1, \dots, l_i \}} f_{i,j}(x) \right).
\end{equation}%
Note that here $f_{i,j}(x)$ is negative if the inequality in \eqref{eq:standard} is satisfied, disjunction is replaced by maximization, and conjunction by minimization. As a result, if $\rho_\varphi(x)$ is negative, $\varphi$ is true and $\rho_\varphi(x)$ is positive otherwise.
Now, based on the measure $\rho_\varphi$, we construct an error metric:
\begin{equation}
e_\varphi(x) := \max( \rho_\varphi(x),0),
\label{eq:errormetric}
\end{equation}
which for a given point $x$ is equal to zero if $\varphi$ is true and positive if not. 
Based on the error metric \eqref{eq:errormetric}, we construct a sample-based fitness over a finite set of samples $\hat{X} = \{x_1, \dots, x_p\} \subset X$ as:
\begin{equation}
\label{eq:sampfit}
\mathcal{F}_{\mathrm{samp},\varphi} := (1+\left\| [ e_\varphi(x_1), \dots, e_\varphi(x_p) ] \right\| )^{-1}.
\end{equation} 
By definition $\mathcal{F}_{\mathrm{samp},\varphi} \in [0,1]$ and is equal to 1 if for all $x \in \hat{X}$ the propositional logic formula $\varphi$ is true. 

Besides sample-based testing, the logic formula is formally verified by means of the SMT solver. Given the output of the SMT solver, the SMT-based fitness is defined as
\begin{equation}
\label{eq:smtfit}
\mathcal{F}_{\mathrm{SMT},\varphi} = \left\{\begin{array}{ll}
1, \text{ if } \neg \varphi~ \text{ is unsat},\\
0, \text{ if } \neg \varphi~ \text{ is }\delta \text{-sat}.
\end{array} \right.
\end{equation}%
Finally, the full fitness of a pair $(V,\kappa)$ satisfying the conditions in Proposition \ref{thm:2} is defined as a weighted sum of the sample-based and SMT-based fitness for each condition. The weighting is motivated by the intuition that prior to checking the conditions of the derivative and the jumps (inequalities \eqref{eq:conVjump}, \eqref{eq:conVd}, \eqref{eq:conVj}, and \eqref{eq:conVjc}), $V$ must first have the `correct shape', i.e. satisfy the conditions with respect to the initial set and safe set (inequalities \eqref{eq:conI} and \eqref{eq:condS}). Therefore, the conditions are sequentially weighted with 
\begin{align*}
w_i = \lfloor w_{i-1} \mathcal{F}_{\mathrm{samp},\varphi_{i-1}} \rfloor,~~~i \in \{2,\dots 6\}, 
\end{align*}
and $w_1=1$, where for each $\varphi_i$ the corresponding inequality is shown in Appendix \ref{app:standard} in Table \ref{tab:sfcon}.
The overall fitness is then defined as:
\begin{equation}
\label{eq:fit}
\mathcal{F}:= \frac{1}{12} \sum_{i=1}^{6} w_i \left( \mathcal{F}_{\mathrm{samp},\varphi_i}+\mathcal{F}_{\mathrm{SMT},\varphi_i} \right) .
\end{equation}
Note that $\mathcal{F} \in [0,1]$ and only if $\mathcal{F} = 1$, all conditions are formally proven by means of the SMT solver, hence the candidate function $V$ is an LBF. In case it is desired to verify conditions from Corollary \ref{col:rsws2}, the fitness function is extended in a similar way. 

\begin{rem}[Robustness w.r.t. $\delta$-sat]
\label{rem:dsatrobustness}
As stated before, $\delta$-sat and unsat are not always mutually exclusive. If both are true, dReal can return either case. To circumvent this overlap, candidate solutions are synthesized such that they are robust w.r.t. the $\delta$ perturbation. This is done by strengthening the inequalities used in the sampled-based fitness relatively to the $\delta$ perturbation. That is, for some $\epsilon \geq \delta$ and a formula expressed as \eqref{eq:standard}, the sample-based fitness is redefined using the following strengthened formula:
\begin{equation*}
\varphi':= \forall x \in X,~ \left( \bconj_{i=1}\nolimits^{k} \left( \bdisj\nolimits_{j=1}^{l_i}f_{i,j}(x) + \epsilon \leq 0 \right) \right).
\end{equation*}
\end{rem}

\subsubsection{Secondary fitness measures}
\label{sec:secfit}
In case two or multiple individuals have the same fitness value, secondary fitness measures are used to rank individuals. The first secondary fitness value is based on the number of parameters and the second secondary fitness is based on the norm of all the parameter values. The latter promotes less complex but equivalent individuals, and the former aims to prevent parameters to blow up without improving the fitness.

\subsection{Algorithm outline}
\label{sec:outline}
Given a system $\mathcal{H}_\mathrm{cl}$, compact sets $(S,I,O)$ and a grammar, the algorithm undergoes the following steps:

\begin{enumerate}
\item A random population of $(V,\kappa)$ tuples is generated adhering to the provided grammar. 
\item The parameters within the structure of each individual are adjusted using CMA-ES to optimize the sample-based fitness.
\item For all individuals with full sample-based fitness, an SMT solver is used. If there is a violation, counterexamples are generated by the SMT solver, which are added to the set employed in the sample-based fitness.
\item The overall fitness in \eqref{eq:fit} is computed for all individuals.
\item A new population is generated by:
\begin{enumerate}
\item Copying the best individuals of the current generation.
\item Selecting individuals using tournament selection and modifying them using genetic operators.
\end{enumerate}
\item Steps 2 to 5 are repeated until the maximum fitness value (i.e. 1) is obtained, or a maximum number of generations is met. 
\end{enumerate}

\begin{table}[t]
\centering
\caption{Continuous-time systems with input $u \in [\underline{u},\overline{u})]$. 1: linear system. 2: 2nd order polynomial system. 3: 3rd order polynomial system. 4: Pendulum system. 5: Pendulum-on-cart system.}
\label{tab:systems}
\scalebox{0.9}{
    \begin{tabular}{cccc}
    \toprule
 System & $f_\mathrm{ol,x}(x,u)$ & $(S_\mathrm{x},I_\mathrm{x},O_\mathrm{x})$ & $(\underline{u},\overline{u})$ \\
 \midrule
1& $\begin{pmatrix}
 x_2 \\ 
 -x_1 +u
\end{pmatrix}$ & 
$\begin{array}{l}
([-1,~ 1]^2,\\ ~[-0.5, 0.5]^2,\\~ [-0.1, 0.1]^2)
\end{array}$
&(-1,1)\\
2 & $\begin{pmatrix}
x_2-x_1^3\\ 
u
\end{pmatrix} $
& 
$\begin{array}{l}
([-1,~ 1]^2,\\~ [-0.5, 0.5]^2, \\~[-0.05,~0.05]^2)
\end{array}$
&
(-1,1)
\\
3 & 
$\begin{pmatrix}
-10x_1 +10 x_2+u \\ 
28 x_1 - x_2-x_1 x_3 \\
 x_1 x_2 - 2.6667 x_3
\end{pmatrix} $ & 
$\begin{array}{l}
([-5,~ 5]^3,\\~[-1.2, 1.2]^3,\\ ~[-0.3, 0.3]^3)
\end{array} $ 
&
(-100,100)\\
4 & 
$\begin{array}{l}
\begin{pmatrix} x_2 \\
 \frac{m l g}{J} \sin(x_1) - \left(\frac{b}{J}+\frac{K^2}{J R_a} \right) x_2 +\frac{K}{J R_a}u
  \end{pmatrix} \\
 m =5.50\cdot 10^ {-2} \textrm{ kg},~ l = 4.20 \cdot 10^ {-2} \textrm{ m},\\
 J = 1.91\cdot 10^{-4}\textrm{ kg} ~ \textrm{m}^2,~ g = 9.81 \textrm{ m} / \textrm{s}^2, \\
 K = 5.36\cdot 10^ {-2} \textrm{ Nm/A},~R_a = 9.50 \Omega. \\ 
 b= 3.0\cdot 10^{-6} \textrm{Nms}
\end{array}$
&
$\begin{array}{l}
([-2 \pi,~ 2\pi] \\
\times \left[-100,~100\right],\\
~[-\pi,~\pi] \\
\times ~[-10,~10], \\
 ~[-1.0,~ -0.5] \\
 \times ~[-1.0,~1.0])
\end{array} $ 
&
(-10,10)\\
5& $\begin{array}{l}
\begin{pmatrix}
x_2\\ 
\frac{g}{l}\sin(x_1) - \frac{b}{ml^2}x_2+\frac{1}{ml}\cos(x_1) u
\end{pmatrix} \\
 g = 9.8 \textrm{ m}/ \textrm{s}^2 ,~ b = 2~\textrm{Nms} \\
 l = 0.5\textrm{ m}, ~ m =0.5\textrm{ kg}.
\end{array}$ 
& 
$\begin{array}{l}
 ([-2\pi,2\pi]\\
 \times [-10,10], \\
 ~[-0.5, 0.5]^2, \\ 
 ~[-0.25, 0.25]^2
\end{array} $&
(-6,6) \\
\bottomrule
  \end{tabular}%
  }
\end{table}
\section{Case studies}
In this section we demonstrate the effectiveness of the proposed approach on several benchmark systems. Here we consider continuous-time systems, sampled-data systems, uncertain systems, switching controllers, and fully hybrid systems. All benchmarks were performed using an Intel Xeon CPU E5-1660 v3 3.00GHz using 14 parallel CPU cores. The GGGP and CMA-ES algorithms were both implemented in Mathematica 11.1. 

Within the synthesis, the choice of the grammar is essential. In this work, we use a grammar covering polynomials and/or use case-specific insights to bias the grammar. Here the use of polynomials is motivated by the Weiestrass approximation theorem, stating that any continuous function on a closed interval can be approximated arbitrarily close by a polynomial. Regardless, there still might not exist a polynomial LBF \cite{Ahmadi2011} or it might yield a very high-order polynomial, such that the use of transcendental functions like sine functions or exponentials might be more beneficial.

\subsection{Continuous open-loop systems}
First of all, we consider fully continuous-time open-loop systems, i.e. with $D_\mathrm{s} = \emptyset$. We consider five systems, adopted from \cite{Verdier2018} and references therein, defined by the open-loop continuous dynamics $f_\mathrm{ol,x}: \R^{n_\mathrm{x}} \times \mathcal{U} \rightarrow \R^{n_\mathrm{x}}$ shown in Table \ref{tab:systems}. These systems are: a linear system, 2nd- and 3rd-order polynomial systems, a pendulum system, and a pendulum-on-cart system. We consider saturated control inputs, i.e. controllers of the form
\begin{align*}
\kappa(x) &= \mathrm{sat}_{(\underline{u},\overline{u})} \circ \kappa'(x),\\
\mathrm{sat}_{(\underline{u},\overline{u})}(x) &= \max(\underline{u}, \min(\overline{u}, x)),
\end{align*}
where $\kappa': \R^{n_\mathrm{x}} \rightarrow \R^m$ is an analytic controller to be synthesized by the proposed framework. Furthermore, we consider the system with continuous full state-feedback and with sampled-data input. In the former case the system dynamics is given by $\mathcal{H}_\mathrm{ct}$ with data
\begin{align*}
C = \R^{n_\mathrm{x}},~F(s) =f_\mathrm{ol,x}(s_\mathrm{x},\kappa \circ h(s)), \\
D = \emptyset,~ G(s) = \emptyset, ~h(s) = s_\mathrm{x}.
\end{align*}
Given a sampling time $\eta>0$, the effect of sampled data can be modeled by adding the sampled states as additional discrete states, resulting in the system $\mathcal{H}_\mathrm{sd}$ with
\begin{align*}
C = \R^{2 n_\mathrm{x}} \times [0,\eta],~F(s) = \begin{pmatrix} f_\mathrm{ol,x}(s_\mathrm{x}, \kappa \circ h(s)), & \mathbf{0}_{n_\mathrm{x}}, & 1 \end{pmatrix}, \\
D = D_\mathrm{t}= \R^{2n_\mathrm{x}} \times \{\eta \},~G(s) = (s_\mathrm{x}, s_\mathrm{x}, 0),~h(s) = s_\mathrm{q}.
\end{align*}
Note that here $h(s)$ is dependent on the discrete states $s_
\mathrm{q}$. Given these models $\mathcal{H}_\mathrm{ct}$ and $\mathcal{H}_\mathrm{sd}$, we synthesize controllers $\kappa'$ and LBFs $V$ for specification \CS{1} with $(S,I,O)$ as $(S_\mathrm{x}, I_\mathrm{x}, O_\mathrm{x})$ for $\mathcal{H}_\mathrm{ct}$ and as $(S_\mathrm{x}^2 \times \mathcal{T},\{ (s_\mathrm{x},s_\mathrm{q},s_\mathrm{t}) \in I_\mathrm{x}^2 \times \{0\} \mid s_\mathrm{q} = s_\mathrm{x}\},O= O_\mathrm{x}^2 \times \mathcal{T} )$ for $\mathcal{H}_\mathrm{sd}$, where
$(S_\mathrm{x}, I_\mathrm{x}, O_\mathrm{x})$ are defined for each system in Table \ref{tab:systems}, and $\eta$ as shown in Table \ref{tab:CTres}.

As a baseline of the proposed framework, we synthesize controllers and LBFs based on parametrized candidate solutions with fixed structures. Since the structure is fixed, no genetic operators are applied. This is a special case of the full framework, where the grammar specifies a single full candidate template.
For these parametrized solutions, we consider for models $\mathcal{H}_\mathrm{ct}$ and $\mathcal{H}_\mathrm{sd}$ templates of the form
\begin{align*}\
(V(s), \kappa'(s))_\mathrm{ct} &= ( x^T A_1 x + c,~ K x),\\
(V(s), \kappa'(s))_\mathrm{sd} &= ( x^T A_1 x +(\eta- s_\mathrm{t})te^T A_2 e +c, K z),
\end{align*}
respectively, where $x = s_\mathrm{x} -x_O$, $z = s_\mathrm{q} -x_O$, $e = x-z$, $A_1, A_2$ are upper-triangular matrices, $c$ a constant, and $x_O$ the center of $O_\mathrm{x}$.
We consider 14 individuals and start with 100 test samples and a maximum of 300 counterexamples, where a first-in-first-out principle is used. We use per iteration 30 CMA-ES generations and we set the maximum number of iterations to 200.
The results are shown in Table \ref{tab:CTres}. Here we observe that for model $\mathcal{H}_\mathrm{sd}$ of system 3 the computation time of the SMT solver surpassed the user-imposed time-out limit of 20 seconds for all individuals in a generation. In this case no counterexamples are generated, nor an answer is provided whether an individual is a solution, hence the algorithm is terminated. For model $\mathcal{H}_\mathrm{sd}$ of system 4 and the given template, we observe that no solutions are found within the maximum number of iterations. Note that this is no guarantee that no solution exists within this solution structure. 
 
\begin{figure}[t]%
\centering
\includegraphics[scale=1]{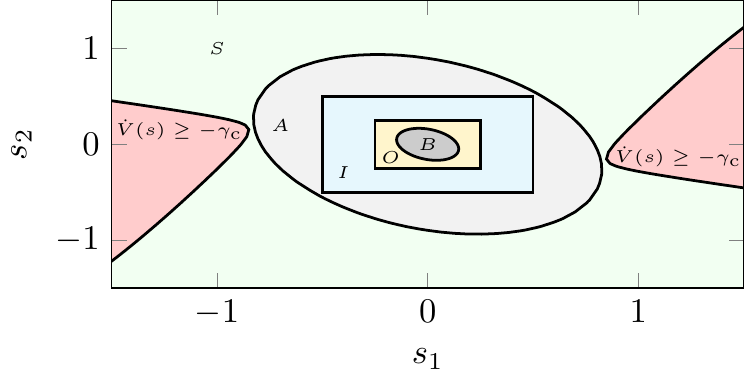}
\caption{Sets $(S,I,O)$ and the sublevel sets $A$ and $B$ of a found result for system 5 with continuous-time controller. The red areas indicate where the derivative $\dot{V}(s) = \langle V(s), F(s) \rangle$ is above $\gamma_\mathrm{c}$.}
\label{fig:pendres}
\end{figure}

Let us consider the solutions for model $\mathcal{H}_\mathrm{ct}$ of system 5. Using a line search over $\beta$ and checking the inequalities in Corollary \ref{col:rsws} using an SMT solver, we found that for 9 out of 10 solutions we could find a $\beta$ such that Corollary \ref{col:rsws} holds, i.e. the closed-loop system also satisfies \CS{2}. An example of a solution that also satisfies \CS{2} is given by
\begin{align*}
V(s) &= -14.4983 + 23.06 s_1^2 + 11.6469 s_1 s_2 + 17.9399 s_2^2, \\
\kappa(s)& = - 11.0776 s_1 - 9.32858 s_2,
\end{align*}
with $\beta = -14.0381$. The sets $S$, $I$, $O$, $A$, and $B$ are shown in Figure \ref{fig:pendres}. Since Corollary \ref{col:rsws} holds, $A$ is a forward invariant set which is found automatically using the proposed framework. Moreover, note that given the found solution, we cannot trivially increase the size of this forward invariant set $A$, e.g. by shifting $V$, as we can observe that for some neighboring states of $A$ we have $\langle \nabla V(s), f(s) \rangle > - \gamma_\mathrm{c}$, which would then violate condition \eqref{eq:conVd}.

\begin{table*}[t]
\caption{Results across 10 runs for continuous-time systems with continuous or sampled-data controllers, using a fixed template. $\mu$: mean, $\sigma$: standard deviation. $~^1$ SMT time-out, ~ $~^2$ No convergence. }
\label{tab:CTres}
\centering
\scalebox{0.72}{  
\begin{tabular}{cccccccccccccccccc}
\toprule
 \multirow{3}{*}{System} & \multicolumn{8}{c}{ continuous-time model $\mathcal{H}_\mathrm{ct}$} & \multirow{3}{*}{$\eta$} & \multicolumn{8}{c}{sampled-data model $\mathcal{H}_\mathrm{sd}$}\\ \cmidrule(lr){2-9} \cmidrule(lr){11-18}
 &  \multicolumn{4}{c}{number of generations} & \multicolumn{4}{c}{time [s]} & &\multicolumn{4}{c}{number of generations} & \multicolumn{4}{c}{time [s]} \\ 
\cmidrule(lr){2-5} \cmidrule(lr){6-9}\cmidrule(lr){11-14} \cmidrule(lr){15-18}
 & $\min$ & $\max$ & $\mu$ & $\sigma$ & $\min$ & $\max$ & $\mu$ &
  $\sigma$& 
  & $\min$ & $\max$ & $\mu$ &
  $\sigma$& $\min$ & $\max$ & $\mu$ &
  $\sigma$\\ 
  \midrule
1 & 1 & 1 & 1.0 & 0.00 & 3.44 & 3.87 & 3.61 & 0.14 & 0.01 & 1 & 7 & 2.7 & 1.83 & 14.49 & 126.28 & 49.82 & 36.00 \\
2 & 1 & 3 & 2.1 & 0.57 & 3.44 & 11.30 & 7.85 & 2.21 & 0.01 & 2 & 6 & 4.1 & 1.37 & 30.33 & 179.11 & 107.35 & 53.13 \\
3 & 2 & 4 & 2.7 & 0.82 & 8.10 & 23.00 & 13.45 & 5.54 & 0.001 & $-^1$ & - &  - &   - &  -  &  -  &  -  &  -  \\
4 & 4 & 9 & 7.0 & 1.76 & 15.90 & 47.29 & 33.63 & 10.95 & 0.001 &  $-^2$ &  - & -  &   - &  -  & -   &  -  &   - \\
5 & 2 & 5 & 2.9 & 0.99 & 7.60 & 23.29 & 12.32 & 5.00 & 0.01 & 3 & 16 & 8.6 & 3.66 & 36.80 & 576.35 & 178.98 & 153.87 \\
\bottomrule
  \end{tabular}%
  }
\end{table*}

\subsection{Bounded uncertainties}
Let us consider a continuous-time system described by $\dot{x}(t) = f(x,d)$, where $d \in \Delta$ is a bounded disturbance and $\Delta$ is compact. This system can be modeled in the framework by writing the dynamics as the set-valued function $F(s) = \{f(s,d) \in \R^n \mid d \in \Delta\}$. As an example, reconsider model $\mathcal{H}_\mathrm{ct}$ of system 5 (pendulum-on-cart) from Table \ref{tab:systems} and adapt $F(s)$ to
\begin{equation*}
F(s) = \left\{f_\mathrm{ol,x}(s_\mathrm{x},\kappa \circ h(s)) + d \mid d \in \Delta \right\},
\end{equation*}
with $\Delta = \{0\}\times[-0.5,0.5]$. Using the same solution template as before, for 10 runs, synthesis took on average 3.3 generations and 24.22 seconds. 
\subsection{Switching controllers}
\label{sec:DCDCsw}
\begin{table}[t]
\centering
\caption{Production rules $\mathcal{P}$.}
\label{tab:PRexamp}
  \begin{tabular}{rl}
  \toprule
  $\mathcal{N}$ & Rules \\ \midrule
$\left< \mathrm{pol} \right> $ & $::=\rmm{const}\times\rmm{mon} ~|~\rmm{pol} + \rmm{pol} $\\
$\left< \mathrm{mon} \right>$ &$::= \rmm{var} ~|~ \rmm{var} \times \rmm{mon} $ \\
 $\left< \mathrm{var} \right> $ & $::=s_1 ~|~\dots ~|~ s_n$ \\
 $\left< \mathrm{const} \right> $ &$::=$ Random Real $\in \left[-10,10\right]$ \\
 \bottomrule
  \end{tabular}%
\end{table}
Using the proposed framework, it is possible to consider switching controllers. Let us consider the DC-DC boost converter system from \cite{Girard2010}. Rewriting this system as a hybrid system (as in Definition \ref{def:hybridsystem} and satisfying Assumption \ref{assump:basicHybrid}), we have $\mathcal{H} = (C,F,\emptyset, \emptyset)$ with:
\begin{align*}
F(s) = \{ A(s) q +b(s) \mid q \in \sigma(\kappa(s)) \},\\
A(s) = 
\begin{pmatrix}
-\frac{s_1}{x_l} \frac{r_0 r_c}{r_0+r_c} -\frac{s_2}{x_l}\frac{r_0}{r_0+r_c}\\
\frac{s_1}{x_c}\frac{r_0}{r_0+r_c}
\end{pmatrix}, 
b(s) = \begin{pmatrix}
-\frac{s_1 r_l}{x_l} + v_s \\
-\frac{s_2}{x_c}\frac{1}{r_0+r_c}
\end{pmatrix},
\end{align*}
where the parameters of the model are as taken in \cite{Girard2010}, $\kappa$ denotes a to be designed state-dependent controller and $\sigma$ is outer semicontinuous switching function defined as
\begin{align}
\sigma(x) &= \left\{ \begin{array}{rl}
1 & \text{if } x > 0, \\
\left[0,1 \right] & \text{if } x = 0, \\
0 & \text {if } x <0 .
\end{array} \right.
\label{eq:sigma}
\end{align}

\begin{figure}[t]%
\centering
\includegraphics[scale=1]{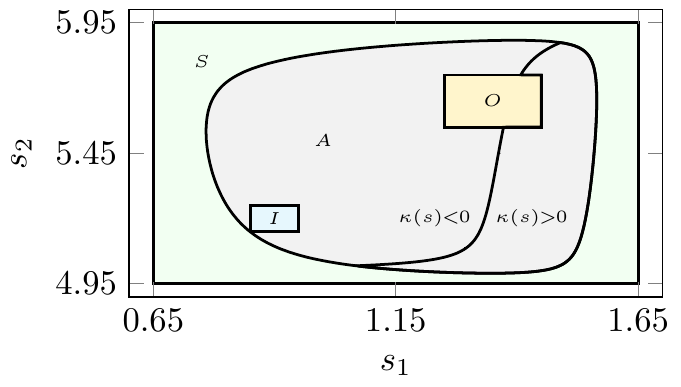}
\caption{Specification set $(S,I,O)$ and the level sets for a found LBF for the DC-DC boost converter system.}
\label{fig:DCDCpipeline}
\end{figure}

We synthesize a controller $\kappa$ for specification \CS{1} with the safe, initial and goal set as in \cite{Ravanbakhsh2017}, i.e. $S = [0.65,1.65]\times[4.95,5.95]$, $I = [0.85, 0.95] \times [5.15,5.25]$, $O=[1.25, 1.45] \times [5.55, 5.75]$. Given that the initial and goal sets are relatively close to the safe set, a second-order polynomial is likely not to suffice, and therefore we bias our solutions by including a pre-specified barrier function of the form:
 \begin{equation*}
B(c,s) = \! \frac{c_1}{1.66\, -s_1}\!+\!\frac{c_2}{5.96\, -s_2}\!+\!\frac{c_3}{s_1-0.64}\!+\!\frac{c_4}{s_2-4.94}.
\end{equation*}
Using this barrier function, we employ the start tree of the candidate LBF $\mathcal{S}_V$ given by the sum of a constant $\rmm{const}$, polynomial $\rmm{pol}$ and the barrier function $B(c)$:
\begin{align*}
\mathcal{S}_V &= \rmm{const} + \rmm{pol} + \rmm{const}B(c,s),\\
c &=\begin{pmatrix}
\rmm{const},& \rmm{const}, &\rmm{const},& \rmm{const}
\end{pmatrix}.
\end{align*}
Let us denote $f_q(s) = A(s)q+b(s)$ for $q \in \{0,1\}$. Taking inspiration from synthesis of switching controllers based on a CLFB (see e.g. \cite{Ravanbakhsh2017}, \cite{Verdier2018}), the controller is based on the candidate LBF $V$, such that 
\begin{align*}
q = \sigma(\kappa(s)) = 1& \text{ if } 
\left< \nabla V(s), f_0(s) \right> > \left< \nabla V(s), f_1(s) \right>,\\
q = \sigma(\kappa(s)) = 0 & \text{ if }
\left< \nabla V(s), f_0(s) \right> < \left< \nabla V(s), f_1(s) \right>.
\end{align*}
In other words, a mode $q$ is selected so that it minimizes $\langle \nabla V(s), f_q(s) \rangle$.
This is achieved by the following controller:
\begin{equation}
\kappa(s) = \left< \nabla V(s), f_0(s) \right> - \left< \nabla V(s), f_1(s))\right>.
\label{eq:dcdcK}
\end{equation}
Based on this prior knowledge, we use the start tree $\mathrm{Tuple}(\mathcal{S}_V, \kappa(s))$ and the production rules in Table \ref{tab:PRexamp}. We used 8 individuals, a maximum tree depth of 10, a mutation chance of 0.8, crossover chance of 0.3, 30 generations in CMA-ES, 100 test samples and a maximum of 300 counterexamples. In 10 different runs with a maximum of 200 generations, we found in 3 runs a solution in the 104th, 115th, and 151st generation with on average 20 seconds per generation. An example of a found solution is given by:
\begin{align*}
V(s) =& 1.66125 B(c,s)-2.96592 s_1^3-5.36934 s_1^2 s_2^2\\
&-26.175 s_1^2 -5.55243 s_1 s_2^2+26.763 s_1 s_2\\
&-17.0781 s_1 -0.0612397 s_2^3 -10.2641 s_2^2\\
&+48.5132 s_2+32.6963,\\
c =& \begin{pmatrix}
28.2706, & 16.4118, &2.64323, & 3.967
\end{pmatrix},
\end{align*}
and the corresponding controller given by \eqref{eq:dcdcK}. Given this solution, the set $A$ and the controller regions are shown in Figure \ref{fig:DCDCpipeline}. While the controller is synthesized for states that start in $I$, specification \CS{1} holds for all states starting in $A$.

\subsection{Discovering controller structures}
In this section we illustrate how our method can be used to automatically find an appropriate controller structure. Here we consider the nonholonomic integrator with $\mathcal{H} = (\R^3, F_\mathrm{ol}(s,\kappa(s)), \emptyset, \emptyset)$ and
\begin{align*}
F_\mathrm{ol}(s,u) =\begin{pmatrix}
u_1,& 
u_2,& s_1 u_2 - s_2 u_1
\end{pmatrix},
\end{align*}
which does not satisfy Brockett's necessary condition \cite{Brockett1993,Schaft2000}. 
Therefore, while this system is controllable, there exists no continuous-time state-feedback law to asymptotically stabilize the system. However, note that this does not automatically imply that there does not exist a continuous state feedback law which satisfies the specifications \CS{1} and \CS{2} for a given $(S,I,O)$. Moreover, we consider a saturated input $u_i =\mathrm{sat}_{(-1,1)} \circ \kappa_i(s)$ for $i \in\{1,2\}$
and a safe set $S =[-5,5]^3$, initial set $I = [-3, 3]^2\times[-0.1,0.1]$ and goal set $O = [-0.5,0.5]^3$. That is, it is desired to steer the system to a neighborhood around the origin, where initially $x_3$ is close to zero.

For simplicity, we consider a parametrized quadratic LBF and for the controller a grammar containing multiple controller classes, namely linear, polynomial, and discontinuous controllers.
The start symbol is given by $\mathcal{S} = \mathrm{Tuple}\left(\rmm{V}, (\rmm{\kappa_i}, \rmm{\kappa_i})\right)$, with
\begin{equation*}
\rmm{V}::= \rmm{const}+\rmm{const}s_1^2+\rmm{const}s_2^2+\rmm{const}s_3^2 ,
\end{equation*}%
and the (other) production rules are given by combining Table \ref{tab:PRexamp} and Table \ref{tab:PRstr2}. In the grammar, $\rmm{disc}$ is the nonterminal for discontinuous expressions and $\sign$ denotes the outer semicontinuous sign function, defined as $\sign(x) := 2 \sigma(x)-1$, where $\sigma$ is defined in \eqref{eq:sigma}.
Finally, the discontinuities are limited to $\sign(s_3)$, to limit the search space and because it is a repeating element in the controllers found in \cite{Schaft2000}. We used 28 individuals, a maximum tree depth of 4, a mutation chance of 0.8, crossover chance of 0.3, a maximum of 200 generations, 30 generations in CMA-ES, 100 test samples and a maximum of 300 counterexamples.

Out of 10 independent runs, the algorithm found in 7 runs a solution within 200 generations. On average, these 7 runs took 19.76 minutes and 110 generations. Of these 7, 6 controllers contained a discrete element in both inputs, 1 controller was fully polynomial, and no linear controllers were found. The polynomial controller is given by
\begin{align*}
V(s) &= -5.3754 + 0.3457 s_1^2 + 0.2184 s_2^2 + 21.6876 s_3^2,\\
\kappa(s) &= \begin{pmatrix}
-0.523878 s_1 + 1.47349 s_2 s_3\\ -0.169653 s_2 - 5.76889 s_1 s_3 + 
 1.16537 s_3^2
\end{pmatrix}.
\end{align*}
Therefore, despite the system not meeting Brockett's necessary condition, for this specification, the algorithm was able to automatically find a sufficient continuous control law, whereas no linear controller was found.

\begin{table}[t]
\centering
\caption{Production rules $\mathcal{P}$ for the nonholonomic integrator controller. }
\label{tab:PRstr2}
  \begin{tabular}{rl}
  \toprule
  $\mathcal{N}$ & Rules \\ \midrule
$\rmm{\kappa_i}$ & $::= \rmm{lin} ~|~ \rmm{pol} ~|~\rmm{pol} + \rmm{const} \rmm{disc}$ \\
$\left< \mathrm{lin} \right> $ & $::= \rmm{const} s_1 + \rmm{const}s_2 + \rmm{const} s_3 $\\ 
$\left< \mathrm{disc} \right> $ & $::= \sign(s_3) ~|~ \rmm{pol}\sign(s_3) ~|$\\ 
 \bottomrule
  \end{tabular}%

\end{table}

\subsection{Jump-flow systems}

\begin{figure}[t]
\centering
\includegraphics[scale =1 ]{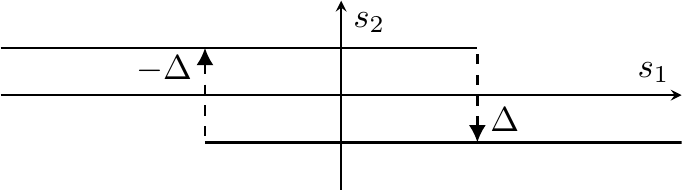}
\caption{Hysteresis.}
\label{fig:hyst}
\end{figure}%
Let us consider a system with $D_\mathrm{s} \neq \emptyset$, namely a hysteresis system adopted from \cite{Branicky1998}, graphically illustrated in Figure \ref{fig:hyst}. This system can be modeled as a hybrid automaton, as shown in \cite{Branicky1998}. Using the jump-flow formalism, the system states $s = (s_\mathrm{x}, s_\mathrm{q}) \in \R \times \{-1,1\}$ consist of a single continuous state $s_\mathrm{x} \in \R$ and a discrete state, which models the state of the hysteresis $s_\mathrm{q} \in \{-1, 1\}$. The system is given by:%
\begin{align*}
F_\mathrm{ol}(s,u) = \begin{pmatrix} s_\mathrm{q} + u, 0 \end{pmatrix}, \\
G_\mathrm{ol}(s,u) = \left\{ \begin{array}{ll} 
\begin{pmatrix} s_\mathrm{x}, -1 \end{pmatrix} & \text{ if } s \in D_1, \\
\begin{pmatrix} s_\mathrm{x} , 1 \end{pmatrix} & \text{ if } s \in D_2, \end{array} \right.\\
H_1 = \{s_\mathrm{x} \in \R \mid s_\mathrm{x} \geq \Delta \}, H_2 = \{s_\mathrm{x} \in \R \mid s_\mathrm{x} \leq -\Delta \},\\
C= [-\Delta,\Delta] \times \{ -1,1\} \cup H_1 \times \{-1\} \cup H_2 \times \{1\}, \\
D_1 = H_1 \times \{1\},~ D_2 = H_2 \times \{-1\}, ~ D_\mathrm{s} = D_1 \cup D_2.
\end{align*}%
Setting $\Delta =1$, we consider the safe, initial and goal set as $(S,I,O) = ([-5,5] \times \{-1 ,1 \}, [-2,2] \times \{-1,1\}, [-1,1] \times\{-0.5,0.5\})$. 
Using the solution template
\begin{align*}
(V(s), \kappa(s)) = ( s A_1 s + c, ~c s_1),
\end{align*}
where $A_1$ is an upper-triangular matrix and $c$ a constant, and using the same settings as before, we synthesized solutions across 10 runs in 2.4 generations and 5 seconds. An example of a solution is given by
\begin{align*}
V(s) =&-228.17+25.027 s_1^2+0.18984 s_1 s_2+84.779 s_2^2,\\
\kappa(s) =&-11.7482 s_1.
\end{align*}

\subsection{Design of flow and jump maps}
Finally, we demonstrate that the approach can also be used to design the flow and jump sets $C$ and $D$. We revisit the DC-DC boost converter from Section \ref{sec:DCDCsw}. Instead of designing a switching signal, we augment the state space with a logic state and design a map $\kappa: \R^2 \rightarrow \R$ that partitions the state space. The closed-loop system is given by the hybrid data $(C,F,D,G)$:
\begin{align*}
F(s) = \begin{pmatrix} A(s_\mathrm{x}) s_\mathrm{q} +b(s_\mathrm{x})\\ 0 \end{pmatrix}, ~
G(s) = \begin{pmatrix} s_\mathrm{x} \\ 1-s_\mathrm{q} \end{pmatrix}, \\
C = \{(x,0) \in S \mid \kappa(x) \leq \varepsilon \} \cap \{(x,1) \in S \mid \kappa(x) \geq 0 \}, \\
D = \{(x,0) \in S \mid \kappa(x) \geq \varepsilon \} \cap \{(x,1) \in S \mid \kappa(x) \leq 0 \}, 
\end{align*}
where $x \in \R^{n_\mathrm{x}}$ and $\varepsilon > 0$. Note that infinite switching between the modes is prevented by design by a hysteresis parametrized by $\varepsilon> 0$. Moreover, as $C \cap D \neq \emptyset$, solutions are not unique, but regardless, the synthesis guarantees that all maximal solutions satisfy the specification. We use again the same expert insight as in Section \ref{sec:DCDCsw} and set $\kappa(x)$ to be equal to the controller structure in \eqref{eq:dcdcK}.
We find that for $\varepsilon = 0.001$, the previously found solution in Section \ref{sec:DCDCsw} is again an LBF.

\section{Discussion and conclusion}
We have proposed a formal synthesis method for hybrid systems. The method is not complete, i.e. solutions may not be found even if they exist. This may stem from a not sufficiently expressive choice of grammar, or due to used optimizers (GGGP and CMA-ES) which do not guarantee finding a solution within a fixed number of generations. While the average computation time in the results of continuous-time systems in Table \ref{tab:CTres} suggests that the computational time increases as the systems become more nonlinear, general conclusions on the computation time are speculative. Besides the absence of convergence guarantees, the computation time can highly vary depending on factors such as system dynamics, system order, grammar, and the GP parameters. Therefore, the method is best used in combination with expert knowledge, incorporated in the grammar. Still, the required expert knowledge is less than to e.g. sum of squares programming or counterexample-guided synthesis approaches, where the user has to provide a solution structure. 

Comparing the method with SCOTS and ROCS for the inverted pendulum system (system 5 in Table \ref{tab:systems}), we obtained a controller in the form of a simple expression, generated after 178.98 seconds on average, whereas according to \cite{Li2018}, with a state grid size of 0.001 the abstraction used in SCOTS took more than 12 hours and did not return a result, and ROCS generated a controller in 400 seconds with a controller consisting of 26340 partitions. Additionally, as the proposed method does not depend on discretization of the state and input spaces, it might yield better scalability w.r.t. memory usage. Compared to the CEGIS methods in \cite{Ravanbakhsh2015,Ravanbakhsh2017}, our method is overall slower, but is able to discover the solution structures itself, whereas e.g. for the DC-DC boost converter the authors of \cite{Ravanbakhsh2017} had to iteratively add barrier functions by hand before a solution was found. Moreover, in our benchmarks we synthesize sampled-data controllers with a larger sampling time than the minimum dwell-times presented in \cite{Ravanbakhsh2015, Ravanbakhsh2017}.

Finally, we are not able to find sampled-data controllers for all systems in Table \ref{tab:systems} (systems 3 and 4) as opposed to our previous work \cite{Verdier2018}. In the case of system 3, this is due to time-out issues with the SMT solver as a result of the increased complexity w.r.t. increased system order. For system 4, the proposed approach is too conservative. Nevertheless, the assumptions on the system dynamics in \cite{Verdier2018} or the need to bound the Lagrangian remainder beforehand are removed in the present framework.

Future and ongoing work focuses on extending the approach to full (signal) temporal logic properties. Similar to e.g. \cite{Kloetzer2008,Ravanbakhsh2017}, this could be done 
by combining B\"uchi automata and our current approach to simple safe reachability.
Finally, more efficient implementations exploiting paralellization, e.g. by using GPU-based computation and more advanced GP variants should improve speed and scalability.

\bibliographystyle{IEEEtran}   
\bibliography{MyBib}

\appendix
\section{Proofs}   
\label{app:proofs}

\subsection{Proof Theorem \ref{thm:1}}
\label{app:thm1proof}
Given that $\phi(0,0) \in I$, if $\phi(0,0) \in O$, \eqref{eq:RWS} holds trivially. For $\phi(0,0) \in I\backslash O$, from \eqref{eq:conI} we have that $V(\phi(0,0)) \leq 0$ and thus $\phi(0,0) \in A^*:= A \backslash O$. From condition \eqref{eq:condS0} and the definition of $A$ in \eqref{eq:defA} we have $A \cap \partial S= \emptyset$. 
Consider a hybrid time interval $[t_j, t_{j+1}]\times \{j\} \subseteq \mathrm{dom} \phi$. For almost all $t\in [t_j,t_{j+1}]$ such that $\phi(t,j) \in A_C^*$, we have from \eqref{eq:conVd} that
\begin{equation*}
    \frac{d}{d t}  V (\phi(t,j)) \leq \max_{f\in F(\phi(t,j))} \left< \nabla V, f \right> \leq - \gamma_\mathrm{c},
\end{equation*}
i.e. $V$ decreases along the flow, hence solutions remain in the sublevel set $A \subset S$ and thus cannot leave the safe set within an arbitrarily small time step. From \eqref{eq:conVjump} it follows that all jumps starting from $A^*_D$ jump to the safe set $S$. 
Moreover, from condition \eqref{eq:conVj0} it follows that for $\phi(t,j) \in A_D^*$ with $(t,j+1) \in \mathrm{dom} \phi$:
\begin{equation*}
    V(\phi(t,j+1)) \leq V(\phi(t,j)) - \gamma_\mathrm{d},
\end{equation*}
i.e. $V$ decreases along a jump, hence solutions remain in the sublevel set $A$. Summarizing, all $\phi(t,j) \in A^*$ remain in $A \subset S \subset C \cup D$ under an arbitrarily small interval of time and/or jump. 

Now, by contradiction, we prove that eventually all trajectories starting in $A^*$ enter $O$. Consider a complete solution which always remains within $A^*$. Since $V(s)$ is continuous and $S$ is compact, $V[S] \subset \R$ is compact and hence $V[S \backslash O] \subseteq V[S]$ is bounded, i.e. $\exists e \in \R$ such that $\forall s \in A^*$, $V(s) \geq e$. 
Using $\forall (t,j) \in \mathrm{dom}\phi: V(\phi(t,j)) \in A^*$, equation \eqref{eq:conVj0}, integrating both sides of \eqref{eq:conVd}, and $V(\phi(0,0)) \leq 0$, we have 
\begin{equation}
V(\phi(t,j)) \leq -t \gamma_\mathrm{c} -j \gamma_\mathrm{d}.
\end{equation}
Since the maximal solution $\phi$ is complete, $j$ is unbounded and/or $t$ is unbounded, which implies in both cases that there exists a finite $T$ and $J$ such that $V(\phi(T,J)) < e$ and thus $\phi(T,J) \notin A^*$, contradicting the premise. Since all $\phi(t,j) \in A^*$ cannot leave $A \subset S \subset C \cup D$ within an arbitrarily small interval of time and/or a number of jumps, the only possibility is that there exists a $(T,J) \in \mathrm{dom} \phi$ such that $\phi(T,J) \in O$, and thus \eqref{eq:RWS} holds. \qed

\subsection{Proof Corollary \ref{col:rsws}}
From Theorem \ref{thm:1} we have that for all maximal solutions $\phi \in \mathcal{S}_{\mathcal{H}_\mathrm{cl}}(I)$, there exists a pair $(T,J) \in \mathrm{dom} \phi$ such that $\phi(T,J) \in O$. Analogous to the proof of Theorem \ref{thm:1}, conditions \eqref{eq:conVjumpadd0}, \eqref{eq:conVd2}, and \eqref{eq:conVj20} imply that $\forall \phi(t,j) \in O$, $\exists (T_1,J_1) \in \mathrm{dom} \phi$ such that $\phi(T_1,J_1) \in B$. 

Since  $B := \{ s \in O \mid V(s) \leq \beta \}$ and $O$ is compact, it follows that $B$ is compact. Condition \eqref{eq:conVd2} implies that 
\begin{equation}
\forall s \in \partial B \cap C, \forall f \in F(s): \langle \nabla V(s), f \rangle \leq  -\gamma_\mathrm{c}.
\end{equation}
Combining this with \eqref{eq:condG0}, we have that all states $\phi(t,j) \in \partial B \cap C$ cannot reach $\partial O$. Therefore it follows that during flow, trajectories starting in $B$ remain within $B\subset O$. From \eqref{eq:conVjumpadd} we have if $ \{(t,j),(t,j+1)\} \subset \mathrm{dom}\phi$ and $\phi(t,j) \in B \cap D$, it follows that $\phi(t,j+1) \in B$, hence for all jumps starting in $B$ the solutions $\phi$ remain within $B \subset O$. Summarizing, solutions within $B$ stay within $B$ and thus $B$ is forward invariant. Since $O \subset S$, we have that \eqref{eq:RSWS} holds.\qed

\subsection{Proof Proposition \ref{thm:2}}

From condition \eqref{eq:condS} and the definition of $A$ in \eqref{eq:defA} we have 
$A \cap (\partial S_\mathrm{x} \ \times S_\mathrm{q} \times \mathcal{T}) = \emptyset$.
During flow, the discrete states $\phi_\mathrm{q}$ remain constant and the timer states $\phi_\mathrm{t}$ remain within $\mathcal{T}$. Therefore, the solution can only escape the safe set $S := S_\mathrm{x} \times S_\mathrm{q} \times \mathcal{T}$ through the boundary of the safe set of continuous states $\partial S_\mathrm{x} \times S_\mathrm{q} \times \mathcal{T}$. Analogous to the proof of Theorem \ref{thm:1}, for $\phi(t,j) \in A_C^*$, $V$ decreases along the flow, and therefore trajectories cannot leave the sublevel set $A$ and thus neither the safe set $S$ within an arbitrarily small time step. Analogous to the proof of Theorem \ref{thm:1}, jumps from $A^*_D$ remain in $A$. 

Now, one can show by contradiction that complete solutions cannot remain forever in $A^*$. Again, $\exists e \in \R$ such that $\forall s \in A^*$, $V(s) \geq e$. Let us denote the number of jumps resulting from $(D_\mathrm{s},G_\mathrm{s})$ and $(D_\mathrm{t},G_\mathrm{t})$ by $j_\mathrm{s}$ and $j_\mathrm{t}$, respectively.
Using $\forall (t,j) \in \mathrm{dom}\phi: V(\phi(t,j)) \in A^*$, equations \eqref{eq:conVj}, \eqref{eq:conVjc}, integrating both sides of \eqref{eq:conVd}, and $V(\phi(0,0)) \leq 0$, yields
\begin{equation}
V(\phi(t,j)) \leq -t \gamma_\mathrm{c} -j_\mathrm{s} \gamma_\mathrm{d}. 
\end{equation}
By the definition of the dynamics of the timer states we have that $j_\mathrm{t}$ depends on time, i.e.: $
j_\mathrm{t}(t) = \sum_{i=1}^{n_\mathrm{t}} \left\lfloor \frac{t+\phi_{\mathrm{t},i}(0,0)}{\eta_i} \right\rfloor
$. Since the maximal solution $\phi$ is complete, $t$ is unbounded and/or $j = j_\mathrm{s} + j_\mathrm{t}$ is unbounded because $t$ is unbounded or $j_\mathrm{s}$ is unbounded. In all cases there exists a finite $T$ and $J$ such that $V(\phi(T,J)) < e$ and thus $\phi(T,J) \notin A^*$. The remainder of the proof is analogous to the proof of Theorem \ref{thm:1}.\qed

\subsection{Proof Corollary \ref{col:rsws2}}

This proof is analogous to the proof of Corollary \ref{col:rsws}, where Proposition \ref{thm:2} is used instead of Theorem \ref{thm:1}. Analogous to condition \eqref{eq:condG0} in Corollary \ref{col:rsws}, condition \eqref{eq:condG} yields that all states $\varphi(t,j) \in \partial B \cup C$ cannot reach $\partial O_\mathrm{x} \times O_\mathrm{q} \times \mathcal{T}$. Since the discrete states $\phi_\mathrm{q}(t,j)$ remain constant during flows and the timer states $\phi_\mathrm{t}(t,j)$ always stay within $\mathcal{T}$, it follows that during flow, trajectories starting in $B$ remain within $B\subset O:= O_\mathrm{x} \times O_\mathrm{q} \times \mathcal{T}$. The remainder of the proof is analogous to Corollary \ref{col:rsws}. \qed

\subsection{Proof Corollary \ref{cor:noZeno2}}
From the proof of Corollary \ref{col:rsws2} it follows that $\forall \phi \in \mathcal{S}_{\mathcal{H}_\mathrm{cl}}(I)$, $\exists (T,J) \in \mathrm{dom} \phi$ such that $\forall (t,j) \in E_{\geq (T,J)}$, $\phi(t,j) \in B$. Since $B \cap D_\mathrm{s} = \emptyset$, the only jumps taking place for $(t,j) \in E_{\geq (T,J)}$ are because $\phi(t,j) \in D_\mathrm{t}$, i.e. due to timer updates. Since every jump induced by a timer state has a fixed minimal dwell-time of $\eta_i$ and there are only a finite number of timer states, it follows that all solutions $\phi \in \mathcal{S}_{\mathcal{H}_\mathrm{cl}}(I)$ are non-Zeno. \qed

\subsection{Proof Corollary \ref{cor:persflow}}
As a consequence of the conditions in Theorem \ref{thm:1}, Proposition \ref{thm:2} or Corollaries \ref{col:rsws} and \ref{col:rsws2}, after a jump $\phi(t,j) \in S \subset C \cup D$ and under Assumption \ref{ass:jumps}, we have that $\phi(t,j) \notin D$. Therefore solutions can only be extended through flow, along which the LBF decreases.
\qed

\section{Standard forms of the inequalities in Proposition \ref{thm:2} and Corollary \ref{col:rsws2}}
\label{app:standard}
The conditions in Proposition \ref{thm:2} and Corollary \ref{col:rsws2} can be written in the standard form \eqref{eq:standard}
as shown in Table \ref{tab:sfcon}. In this table $c> 0$ is an arbitrary positive constant, used to cast strict inequalities to non-strict inequalities. Here inequalities over a sublevel set $L_X^a(V) := \{ x \in X \mid V(x) \leq a\}$, e.g. $\forall x \in L_X^c: f(x)\leq 0$, are reformulated by using the logical implication
$\forall x \in X: V(x) \leq a \implies f(x)\leq 0$,
which is equivalent to $\forall x \in X: V(x) > a \vee f(x)\leq 0$.

\begin{table}[t]
\centering
\caption{Standard form \eqref{eq:standard} of the LBF conditions in Proposition \ref{thm:2} and Corollary \ref{col:rsws2}.}
\label{tab:sfcon}
\scalebox{1}{
    \begin{tabular}{cccl}
    \toprule
$\varphi_i$ &eq. & $X$ & $f_{i,j}(x)$ \\
\midrule
\rowcolor[gray]{.9} $\varphi_1$&\eqref{eq:conI} & $I$ & $f_{1,1}(x) = V(x)$.\\
$\varphi_2$ &\eqref{eq:condS} & $\partial S_\mathrm{x} \times S_\mathrm{q} \times \mathcal{T}$ & $f_{1,1}(x) = -V(x)+c$. \\
\rowcolor[gray]{.9} $\varphi_3$& \eqref{eq:conVjump} & $\{(x_1,x_2) \in  (S \backslash O \cap D) \times \R^n \mid x_2 \in G(x_1) \}$ & $f_{i,1}(x) = -V(x_1)+c,$\\ 
\rowcolor[gray]{.9} & & \hfill & $f_{i,2}(x) = b_{S,i}(x_2),~ i \in \{1,\dots, i_S \}.$\\
$\varphi_4$& \eqref{eq:conVd}& $\{(x_1, x_2) \in (S\backslash O \cap C) \times \R^n \mid x_2 \in F(x_1)\}$ & $f_{1,1}(x) = -V(x_1)+c,$ \\ 
& & \hfill & $f_{1,2}(x) = \langle \nabla V(x_1) , x_2 \rangle + \gamma_\mathrm{c}.$\\
\rowcolor[gray]{.9} $\varphi_5$& \eqref{eq:conVj} & $\{ (x_1, x_2) \in (S\backslash O \cap D_\mathrm{s})\times \R^n \mid  x_2 \in G_\mathrm{s}(x_1)\}$ & $f_{1,1}(x_1) = -V(x_1)+c$ 
\\ 
\rowcolor[gray]{.9} & & \hfill & $f_{1,2}(x) = V(x_2) - V(x_1) + \gamma_\mathrm{d}.$\\
$\varphi_6$& \eqref{eq:conVjc} & $\{ (x_1, x_2) \in (S\backslash O \cap D_\mathrm{t}) \times \R^n \mid x_2 \in G_\mathrm{t}(x_1) \}$ & $f_{1,1}(x) = -V(x_1)+c, $ \\ 
& & \hfill & $f_{1,2}(x) = V(x_2) - V(x_1).$ \\
\rowcolor[gray]{.9} $\varphi_7$& \eqref{eq:conVjumpadd0} & $\{ (x_1, x_2) \in (O \cap D)\times \R^n \mid x_2 \in G(x_1) \}$ & $f_{i,1}(x) = V(x_1)-\beta +c,$ \\
\rowcolor[gray]{.9} & & \hfill & $ f_{i,1}(x) = b_{S,i}(x_2),~ i = \{1,\dots, i_S \}.$\\
$\varphi_8$& \eqref{eq:conVd2}& $\{ (x_1, x_2) \in (O \cap C) \times \R^n \mid x_2 \in F(x_1) \}$ & $f_{1,1}(x) = V(x_1)-\beta +c,$ \\
& & \hfill  & $f_{1,2}(x) = \langle \nabla V(x_1) , x_2 \rangle + \gamma_\mathrm{c}.$ \\
\rowcolor[gray]{.9} $\varphi_9$& \eqref{eq:conVj2} & $\{ (x_1, x_2) \in (O \cap D_\mathrm{s})\times \R^n \mid  x_2 \in G_\mathrm{s}(x_1) \}$ & $f_{1,1}(x) = V(x_1)-\beta +c,$ \\ 
\rowcolor[gray]{.9} & & \hfill & $f_{1,2}(x) = V(x_2) - V(x_1) + \gamma_\mathrm{d}.$ \\
$\varphi_{10}$& \eqref{eq:conVj2c} & $ \{ (x_1, x_2) \in (O \cap D_\mathrm{t}) \times \R^n \mid x_2 \in G_\mathrm{t}(x_1) \}$ & $f_{1,1}(x) = V(x_1)-\beta +c,$ \\
& &  \hfill  & $f_{1,2}(x) = V(x_2) - V(x_1).$\\
\rowcolor[gray]{.9} $\varphi_{11}$& \eqref{eq:condG} & $\partial O_\mathrm{x} \times O_\mathrm{q} \times \mathcal{T}$  & $f_{1,1}(x) = -V(x)+\beta +c,$.\\
$\varphi_{12}$& \eqref{eq:conVjumpadd} & $\{ (x_1,x_2) \in (O \cap D) \times \R^n \mid x_2 \in G(x_1) \}$ & $f_{i,1}(x) =-V(x_1)+\beta +c,$\\
& & \hfill  & $f_{1,2}(x) = V(x_2)-\beta,$ \\
& & & $f_{k+1,2}(x) = b_{O,k}(x_2),$ \\
& & & $k \in \{1,\dots, i_O\},~ i \in \{1,\dots i_O +1\}. $ \\
    \bottomrule
    \end{tabular}%

    }

\end{table}

\section{List of symbols}
\label{app:LOS}
\textbf{Hybrid systems}
\begin{labeling}{$\mathcal{S}_\mathcal{H}(I)$~~~~ }
\item [$C, D$] Flow and jump set
\item [$F,G$] Flow and jump map
\item[$\mathcal{H}$] Hybrid system 
\item[$E$] Hybrid time domain 
\item [$E_{\leq (T,J)}$]$E_{\leq (T,J)} := E \cap ( [0,T] \times [0,J])$
\item[$E_{\geq (T,J)}$] $E_{\geq (T,J)} := E\backslash ( [0,T) \times [0,J))$
\item[$\phi$ ] Hybrid arc / solution to a hybrid system 
\item[$\mathcal{S}_\mathcal{H}(I)$ ] Set of all maximal solutions starting from $I$ 
\end{labeling}
\textbf{Problem definition} 
\begin{labeling}{$S,I,O$~~}
\item[$F_\mathrm{ol}, G_\mathrm{ol}$] Open-loop flow and jump map
\item [$\mathcal{H}_\mathrm{cl}$] Closed-loop hybrid system
\item [$\kappa, h$] Controller and output map 
\item [$S,I,O$] Safe, initial and goal sets
\end{labeling}
\textbf{Lyapunov barrier function}
\begin{labeling}{$A, A^*,A_Y^*$ ~}
\item[$V$] Lyapunov barrier function
\item[$A, A^*$ ] $A:= \{s \in S \mid V(s) \leq 0\}$, $A^* := A \backslash O$
\item[$A_Y^*$]$A_Y^*:= (A \backslash O) \cap Y$
\item[$B$] $B:= \{ s \in O \mid V(s) \leq \beta \}$
\item[$O^*,O^*_Y$ ] $O^* = O \backslash int(B)$, $O_Y^*:= O^* \cap Y$
\end{labeling}
\textbf{Relaxations}
\begin{labeling}{$G_\mathrm{ol,s}, G_\mathrm{ol,t}$~}
\item[$\phi_\mathrm{x}, \phi_\mathrm{q}, \phi_\mathrm{t}$] Continuous, discrete and timer states
\item[$\mathcal{X},\mathcal{Q}, \mathcal{T}$] Continuous, discrete and timer states space 
\item [$\eta, \mathrm{reset}$ ] Reset time and timer reset map
\item [$D_\mathrm{s}, D_\mathrm{t}$ ] System and timer jump sets 
\item [$G_\mathrm{ol,s}, G_\mathrm{ol,t}$ ] System and timer jump maps of the open loop
\item[$G_\mathrm{s}, G_\mathrm{t}$ ] System and timer jump maps of the closed loop
\item [$S_\mathrm{x},I_\mathrm{x},O_\mathrm{x}$ ] Safe, initial and goal sets of the cont. states 
\item[$S_\mathrm{q},O_\mathrm{q}$ ] Safe and initial sets of the discrete states 
\end{labeling}
\textbf{Genetic programming}
\begin{labeling}{$\mathcal{F}_{\mathrm{samp},\varphi}$~~~}
\item[$\mathcal{N},\mathcal{S},\mathcal{P}$] Nonterminals, start symbol and production rules 
\item[$\varphi$ ] First-order propositional logic formula
\item[$\rho_\varphi, e_\varphi$ ] Satisfaction measure and error metric of $\varphi$ 
\item[$\mathcal{F}_{\mathrm{samp},\varphi}$] Sample-based fitness of $\varphi$ 
\item[$\mathcal{F}_{\mathrm{SMT},\varphi}$] SMT-based fitness of $\varphi$
\item[$\mathcal{F}$] Overall fitness 
\end{labeling}

\end{document}